\documentstyle[11pt]{article}
\def\one{1\hskip -.37em 1}     
\def\proj{E\hskip -.67em I}     
\begin{document}
\begin{titlepage}
\begin{centering}

{\ }\vspace{3cm}

{\Large\bf Solving Gauge Invariant Systems without Gauge Fixing\,:}\\
\vspace{0.5cm}
{\Large\bf the Physical Projector in 0+1 Dimensional Theories}\\

\vspace{1.5cm}

Jan Govaerts$^{*}$ and John R. Klauder$^{\dagger}$

\vspace{0.5cm}

$^{*}${\em Institut de Physique Nucl\'eaire}\\
{\em Universit\'e catholique de Louvain}\\
{\em B-1348 Louvain-la-Neuve, Belgium}\\
{\tt govaerts@fynu.ucl.ac.be}\\

\vspace{1cm}

$^{\dagger}${\em Departments of Physics and Mathematics}\\
{\em University of Florida}\\
{\em Gainesville, Florida 32611, USA}\\
{\tt klauder@phys.ufl.edu}\\

\vspace{2cm}

\begin{abstract}

\noindent The projector onto gauge invariant physical states was recently
constructed for arbitrary constrained systems. This approach, which does
not require gauge fixing nor any additional degrees of freedom beyond the
original ones---two characteristic features of all other available 
methods for quantising constrained dynamics---is put to work in the context 
of a general class of quantum mechanical gauge invariant systems. The cases
of SO(2) and SO(3) gauge groups are con\-si\-de\-red specifically, and 
a comprehensive understanding of the corresponding physical spectra is
achieved in a straightforward manner, using only standard methods of
coherent states and group theory which are directly amenable to 
generalisation to other Lie algebras. Results extend by far the few 
examples available in the literature from much more subtle and 
delicate analyses implying gauge fixing and the characterization 
of modular space.

\end{abstract}

\end{centering} 

\vspace{2cm}

\noindent hep-th/9809119\\
September 1998

\end{titlepage}

\setcounter{footnote}{0}

\section{Introduction}
\label{Sect1}

There is no need to emphasize the importance of the gauge
invariance concept in physics and mathematics today, both in the
classical as well as in the quantum realms. Ever since Dirac's
classification of constraints in terms of first- and second-class 
ones\cite{Dirac1}, the former associated to the generators of local 
gauge symmetries, quite a number of quantisation methods have been 
de\-ve\-lo\-ped 
for such systems\cite{Reviews}, culminating with the so-called BFV-BRST 
Hamiltonian or Lagrangian approaches\cite{BFV}, both in the operatorial 
and in the path-integral frameworks. 

Nevertheless, in spite of all the mathematical elegance and profound 
insight offered by these methods, it may seem somewhat odd that in order
to circumvent the unwanted effects of unphysical gauge variant negative
norm quantum states present in a manifestly spacetime covariant dynamics, 
which is what the BFV-BRST approaches are tailored to achieve, unavoidably 
quite a heavy machinery of ghosts and ghosts for ghosts has to be introduced
generically. This is in sharp contrast with Dirac's own approach to 
covariant quantisation, in which physical gauge invariant states are 
simply those which are annihilated by the quantum gauge generators,
without the need to introduce any further degrees of freedom beyond
the original ones.

An additional issue usually not properly addressed within the BFV-BRST
approaches is that of the required gauge fixing of the gauge freedom,
with the ensuing possible Gribov problems\cite{Gribov1}, either of a 
local or of a global character (or both) on the space of gauge orbits 
of the system\footnote{For a detailed discussion, see for example pp. 143-153 
and Chapter 4 of Ref.\cite{Gov1}.}. In fact, whatever the gauge fixing 
procedure, this issue of possible Gribov problems must certainly be 
considered, if only to conclude that it is absent, which is typically not 
the case. Indeed, the BFV-BRST approaches cannot guarantee the absence of 
such Gribov ambiguities given a specific gauge fixing procedure,
while a correct description of the physics of non perturbative effects can
be achieved certainly only for a gauge fixing free of any Gribov ambiguity.

Recently\cite{Klauder1}, the projector for physical gauge invariant states was
constructed for general constrained systems\footnote{Such a physical
projector had already been introduced\cite{Shabanov1} in the case of some
simple gauge invariant systems\cite{Shabanov2}, similar to those analysed
in the present paper.}, within the coherent
state approach to canonical quantisation\cite{Klauder2}. 
This construction proceeds within
Dirac's original covariant quantisation of such systems, without
the need either of additional degrees of freedom, nor of any gauge
fixing procedure. Subsequently, the phy\-si\-cal projector approach was 
shown\cite{Gov2} to be generally free of any Gribov ambiguities, and 
to lead naturally to the correct integration over the space of gauge 
orbits of the system.

To explore further the advantages of the simplicity afforded through
the physical projection operator and coherent states, 
it is worthwhile to show how that approach also allows for an efficient 
resolution of some gauge invariant quantised dynamics. The systems 
considered in the present paper are matter coupled Yang-Mills theories 
in 0+1 dimensions, thus corresponding to quantum mechanical systems rather 
than quantum field theories. Nevertheless, their dynamics remains 
both simple and rich enough to allow for explicit solutions which 
still demonstrate the advantages of the physical projector approach over 
more established ones. In a certain sense, these models may also
be viewed as deriving from the dimensional reduction to time dependent
configurations only, of pure Yang-Mills theories in higher dimensions.
In this respect, it is intriguing to remark also that similar 
compactifications have recently become of much interest in the context 
of dualities and the non perturbative dynamics of M-theory\cite{Mtheory}, 
thus obviously offering a whole new field to which the present methods 
could be applied, including fermionic degrees of freedom.

More specifically, systems with SO(2) and SO(3) local gauge invariance
will be considered presently. In the first instance, both the physical 
spectrum as well as the wave functions for all gauge invariant states will be
constructed explicitly. In the SO(3) case, only the physical spectrum
is derived, leaving further issues to be explored in the general context
of this class of models for any simple compact Lie algebra.
Quite accidently, some of the models considered here have recently been
analysed in a more traditional setting, addressing both the issues
of the necessary gauge fixing and the ensuing possible Gribov problems
very carefully and properly\cite{Pause}. The physical spectra and wave 
functions determined in that work enable a direct comparison with 
our results, and where they may be compared, they do indeed agree. 
However, our approach is simpler and more straightforward, in that 
it applies well known techniques of coherent states, integration over 
group manifolds and representation theory, while that of Ref.\cite{Pause}
must dissect with great care the structure of the space of gauge orbits 
of the system and the explicit resolution of Schr\"odinger's equation
on that space which possesses typical conical singularities.

The paper is organised as follows. First, an SO(2) gauge invariant model
is considered, by providing its general motivations (Sect.\ref{Sect2}),
considering its classical dynamics (Sect.\ref{Sect3}), its canonical
quantisation (Sect.\ref{Sect4}), solving its spectrum of gauge invariant
physical states (Sect.\ref{Sect5}), accounting for the reasons of the
degeneracies of that spectrum (Sect.\ref{Sect6}), and finally constructing
the generating function of the wave functions of its physical states
(Sect.\ref{Sect7}). Similar considerations are then partly developed
in the case of a model with non abelian SO(3) gauge symmetry,
introduced in Sect.\ref{Sect8} and whose physical spectrum is
determined in Sect.\ref{Sect9}, but leaving the detailed investigation
of the wave functions of these states for later work. Finally,
some conclusions and prospects for further developments are presented
in Sect.\ref{Sect10}, while specific results of use in the body of
the paper are provided in two Appendices.

\section{The SO(2) Gauge Invariant Model}
\label{Sect2}

The degrees of freedom of the model are a real variable $\lambda(t)$
and a collection of $2d$ real variables $q^a_i(t)$ $(a=1,2;i=1,2,\cdots,d)$,
whose dynamics is determined from the Lagrangian function\footnote{The
summation convention over repeated indices is implicit throughout, including
in squared quantities.}
\begin{equation}
L=\frac{1}{2g^2}\left[\dot{q}^a_i-\lambda\epsilon^{ab}q^b_i\right]^2-V(q^a_i)
\ \ \ ,
\label{eq:LSO(2)}
\end{equation}
with
\begin{equation}
V(q^a_i)=\frac{1}{2}\omega^2\,q^a_iq^a_i\ \ \ .
\label{eq:V}
\end{equation}
Here, $g$ and $\omega$ are arbitrary parameters with appropriate physical
dimensions, and $\epsilon^{ab}$ is the two dimensional antisymmetric
tensor such that $\epsilon^{12}=+1$.

Clearly, the model is that of a collection of $d$ real ``scalar fields"
$q^a_i$ in 0+1 dimensions, in the defining two dimensional representation 
of SO(2), or rather such a collection of $d$ non relativistic particles
propagating in a two dimensional space with a mass normalised to unity. 
These particles are coupled to the single time component $\lambda$ of 
the SO(2) gauge field, while their coordinates have been rescaled by the gauge
coupling constant $g$. The kinetic term is the square of the associated
gauge covariant time derivative, while $V(q^a_i)$ could be any SO(2) 
invariant potential of the scalar fields, namely invariant under 
rotations in the index $a=1,2$. Quite obviously, the generator of
SO(2) gauge transformations is the total two dimensional angular momentum 
of the system of $d$ particles, which must thus vanish identically at 
all times for gauge invariant configurations. 

Since the gauge field $\lambda$ in 0+1 dimensions may be gauged away, 
these $d$ particles interact with one another through the potential 
$V(q^a_i)$. The choice $V(q^a_i)=0$ would correspond to those particles
being free. However, we choose to work rather with the harmonic potential 
given in (\ref{eq:V}), since the system then not only becomes readily 
tractable as a collection of $d$ spherical harmonic oscillators, 
but also provides for a natural regularisation of its mass spectrum 
which then gains also a mass gap as in any {\em bona fide\/} 
confining gauge theory. Indeed, the dimensional reduction to 0+1 dimensions 
of a pure {\em non abelian\/} Yang-Mills theory leads to a Lagrangian 
generalising that in (\ref{eq:LSO(2)}), with (the generalisation of) $\lambda$
then corresponding to the time component of the gauge field and $q^a_i$ 
to its space components, while the potential $V(q^a_i)$ is then {\em quartic\/}
in these components in the case of a non abelian gauge 
group\footnote{Note that dimensional reduction of a 1+1 dimensional theory 
does not lead to such a quartic potential\cite{Gov2}.}. Since these non abelian 
interactions are presumed to be precisely those responsible for a finite 
mass gap and confinement, introducing a quadratic potential in the abelian 
SO(2) case above does go some way in modeling the non abelian case. 
Likewise for dimensionally reduced non abelian models, introducing 
a quadratic gauge invariant potential as in (\ref{eq:V}) turns the reduced 
model into a collection of anharmonic oscillators with an angular 
frequency $\omega$ as a free regularisation parameter, while nevertheless, 
the quartic term may be removed altogether without spoiling the non 
abelian gauge symmetry of the reduced system.

The type of model defined by (\ref{eq:LSO(2)}) has been considered
previously in the literature\cite{Shabanov1,Shabanov2,Gov2}, 
and quite recently again in Ref.\cite{Pause}. In particular, the latter work
addressed specifically the SO(2) and SO(3) gauge invariant models for 
$d=2$ particles, and solved their gauge invariant physical spectra having also
chosen the harmonic potential (\ref{eq:V}). Consequently in the case $d=2$, our
results are directly comparable, even though the methods are completely
different. Indeed, Ref.\cite{Pause} carefully determines the modular
space of gauge orbits of the system, and quantises it by solving the
Schr\"odinger equation on that space, which thus possesses conical
singularities. In contradistinction, ours is a method which relies
on well established techniques of coherent states and group theory,
without having to address the subtle issues of gauge fixing and possible
Gribov ambiguities. Exploring further the advantages of the present approach
is certainly a worthwhile programme, including higher dimensional
gauge invariant theories.

\section{The SO(2) Model: Classical Analysis}
\label{Sect3}

The SO(2) gauge transformations leaving the Lagrangian (\ref{eq:LSO(2)})
invariant, are given by
\begin{equation}
\lambda'(t)=\lambda(t)\,+\,\dot{\theta}(t)\ \ \ ,\ \ \ 
{q'}^a_i(t)=U^{ab}\left(\theta(t)\right)\,q^b_i(t)\ \ \ ,
\label{eq:gaugetransfSO(2)}
\end{equation}
where $\theta(t)$ is an arbitrary time dependent 
function\footnote{The variations (\ref{eq:gaugetransfSO(2)})
do not induce a surface term in the Lagrangian (\ref{eq:LSO(2)}).},
while $U^{ab}\left(\theta(t)\right)$ is the rotation matrix
\begin{equation}
\left[U^{ab}\left(\theta(t)\right)\right]=
\left(\begin{array}{c c}
	\cos\theta(t) & \sin\theta(t) \\
	-\sin\theta(t) & \cos\theta(t) 
	\end{array}\right)\ \ \ .
\label{eq:Uab}
\end{equation}
Indeed, under (\ref{eq:gaugetransfSO(2)}), the covariant time derivatives
$(\dot{q}^a_i-\lambda\epsilon^{ab}q^b_i)$ transform in the same manner
as do the variables $q^a_i$.

The Euler-Lagrange equations of motion for the particle positions $q^a_i$
are simply
\begin{equation}
\frac{d}{dt}\left[\dot{q}^a_i-\lambda\epsilon^{ab}q^b_i\right]\,-\,
\lambda\epsilon^{ab}\left[\dot{q}^b_i-\lambda\epsilon^{bc}q^c_i\right]=
-g^2\omega^2q^a_i\ \ ,\ \ a=1,2;\ i=1,2,\cdots,d\ \ ,
\end{equation}
while that for the gauge degree of freedom $\lambda$ is the constraint
\begin{equation}
\epsilon^{ab}q^a_i\left[\dot{q}^b_i-\lambda\epsilon^{bc}q^c_i\right]=0\ \ \ ,
\end{equation}
corresponding to an identically vanishing total angular momentum of the
system of $d$ particles.

Clearly, given the gauge transformations (\ref{eq:gaugetransfSO(2)}),
any configuration for $\lambda(t)$ may be gauged away through a
transformation with parameter $\theta(t)$ such that
\begin{equation}
\theta(t)=\theta_0-\int_{t_0}^t dt'\,\lambda(t')\ \ \ ,
\end{equation}
where $\theta_0$ is the value of $\theta(t)$ at $t=t_0$, an arbitrary
integration constant. Consequently, in the gauge $\lambda(t)=0$,
the equations of motion reduce to those of a collection of $d$
harmonic oscillators in two dimensions,
\begin{equation}
\ddot{q}^a_i=-g^2\omega^2q^a_i\ \ \ ,
\end{equation}
constrained to have a vanishing total angular momentum,
$\epsilon^{ab}q^a_i\dot{q}^b_i=0$.

The Hamiltonian analysis of constraints proceeds as in the usual Yang-Mills
case\cite{Reviews}. The ``fundamental first-order Hamiltonian 
description"\footnote{This notion is introduced on pp. 124-134
of Ref.\cite{Gov1}.} is then provided by the first-order Lagrangian
\begin{equation}
L=\dot{q}^a_ip^a_i-\frac{1}{2}g^2p^a_ip^a_i-\frac{1}{2}\omega^2q^a_iq^a_i
-\lambda\epsilon^{ab}p^a_iq^b_i=\dot{q}^a_ip^a_i-H_0-\lambda\phi\ \ \ ,
\end{equation}
where $p^a_i$ are the momenta conjugate to the coordinates $q^a_i$,
whose Poisson brackets with the latter are canonical, while $\lambda$
turns out to be the Lagrange multiplier for the first-class constraint
$\phi$ generating the local SO(2) gauge invariance of the system,
\begin{equation}
\phi=\epsilon^{ab}p^a_iq^b_i\ \ \ ,
\end{equation}
which commutes with the first-class Hamiltonian 
$H_0=\left[g^2(p^a_i)^2+\omega^2(q^a_i)^2\right]/2$,
namely $\{H_0,\phi\}=0$.

Infinitesimal Hamiltonian gauge transformations generated by $\phi$
are thus defined by
\begin{equation}
\delta_\eta q^a_i=\{q^a_i,\eta\phi\}=\eta\epsilon^{ab}q^b_i\ \ ,\ \ 
\delta_\eta p^a_i=\{p^a_i,\eta\phi\}=\eta\epsilon^{ab}p^b_i\ \ ,\ \ 
\delta_\eta\lambda=\dot{\eta}\ \ ,
\end{equation}
where $\eta(t)$ is an arbitrary (infinitesimal) function.
Clearly, these transformations exponentiate to 
finite SO(2) gauge transformations
acting on the Hamiltonian variables as follows,
\begin{equation}
{q'}^a_i(t)=U^{ab}\left(\theta(t)\right)q^b_i(t)\ \ ,\ \ 
{p'}^a_i(t)=U^{ab}\left(\theta(t)\right)p^b_i(t)\ \ ,\ \ 
\lambda'(t)=\lambda(t)+\dot{\theta}(t)\ \ ,
\end{equation}
where $\theta(t)$ is an arbitrary rotation angle, and the matrix
$U^{ab}\left(\theta(t)\right)$ is given in (\ref{eq:Uab}).
These transformations thus coincide with the SO(2) gauge transformations
in the Lagrangian formulation of the system.

The Hamiltonian equations of motion are simply\footnote{Note how the terms
linear in $\lambda$ in these expressions show how the Lagrange multiplier
is indeed related to a SO(2) rotation added onto the straightforward
time evolution generated by the time derivative.}
\begin{equation}
\dot{q}^a_i=g^2p^a_i+\lambda\epsilon^{ab}q^b_i\ \ ,\ \ 
\dot{p}^a_i=-\omega^2q^a_i+\lambda\epsilon^{ab}p^b_i\ \ ,
\end{equation}
together with the gauge constraint $\phi=\epsilon^{ab}p^a_iq^b_i=0$.
Here again, the local SO(2) gauge freedom enables one to gauge away the
Lagrange multiplier $\lambda$, thereby leading in the $\lambda=0$ gauge
to the usual Hamiltonian equations of motion for harmonic oscillators
but constrained to have a vanishing total angular momentum $\phi=0$.

\section{The SO(2) Model: Canonical Quantisation}
\label{Sect4}

Canonical quantisation of the model is straightforward, through the
canonical commutation relations,
\begin{equation}
\left(\hat{q}^a_i\right)^\dagger=\hat{q}^a_i\ \ ,\ \ 
\left(\hat{p}^a_i\right)^\dagger=\hat{p}^a_i\ \ ,\ \ 
\left[\hat{q}^a_i,\hat{p}^b_j\right]=i\hbar\delta^{ab}\delta_{ij}\ \ ,
\end{equation}
and the quantum Hamiltonian and gauge generator,
\begin{equation}
\hat{H}=\hat{H}_0+\lambda(t)\hat{\phi}\ \ ,\ \ 
\hat{H}_0=\frac{1}{2}g^2\hat{p}^a_i\hat{p}^a_i+
\frac{1}{2}\omega^2\hat{q}^a_i\hat{q}^a_i\ \ ,\ \ 
\hat{\phi}=\epsilon^{ab}\hat{p}^a_i\hat{q}^b_i\ \ ,
\end{equation}
which obviously commute also at the quantum level, 
$[\hat{H}_0,\hat{\phi}]=0$, thus establishing
gauge invariance of the quantised system as well.

As usual (see Appendix A), it is convenient to introduce creation and
annihilation operators in these cartesian coordinates, defined by
\begin{equation}
\alpha^a_i=\sqrt{\frac{\omega}{2\hbar g}}\left[\hat{q}^a_i+
i\frac{g}{\omega}\hat{p}^a_i\right]\ \ ,\ \ 
{\alpha^a_i}^\dagger=\sqrt{\frac{\omega}{2\hbar g}}\left[\hat{q}^a_i-
i\frac{g}{\omega}\hat{p}^a_i\right]\ \ ,
\end{equation}
such that
\begin{equation}
\left[\alpha^a_i,{\alpha^b_j}^\dagger\right]=\delta^{ab}\delta_{ij}\ \ \ ,
\end{equation}
while in terms of normal ordered expressions, and the excitation level
operator $\hat{N}={\alpha^a_i}^\dagger\alpha^a_i$,
\begin{equation}
\hat{H}_0=\hbar g\omega\left[{\alpha^a_i}^\dagger\alpha^a_i+\frac{1}{2}2d\right]
=\hbar g\omega\left[\hat{N}+d\right]\ \ ,\ \ 
\hat{\phi}=i\hbar\epsilon^{ab}{\alpha^a_i}^\dagger\alpha^b_i\ \ .
\label{eq:Nhat}
\end{equation}
Related to these definitions as well as to the phase space
degrees of freedom $q^a_i$ and $p^a_i$, the following complex variables
may also be considered,
\begin{equation}
z^a_i=\sqrt{\frac{\omega}{2\hbar g}}\left[q^a_i+
i\frac{g}{\omega}p^a_i\right]\ \ ,\ \ 
\bar{z}^a_i=\sqrt{\frac{\omega}{2\hbar g}}\left[q^a_i-
i\frac{g}{\omega}p^a_i\right]\ \ .
\end{equation}

Given this cartesian basis of degrees of freedom, it is possible
to introduce generating vectors spanning the whole representation space
of the quantised system, namely Fock states as well as holomorphic or
phase space coherent states (see Appendix A). However, as discussed
in Appendix A, in the case of a two dimensional harmonic oscillator with
circular symmetry, it proves efficient to rather work in 
a helicity-like basis, taking thus advantage of the corresponding SO(2)=U(1)
symmetry, which in the present instance coincides also with the
local gauge invariance of the system.

In terms of the cartesian quantities introduced above, the helicity
annihilation and creation operators are thus defined by
\begin{equation}
\alpha^\pm_i=\frac{1}{\sqrt{2}}\left[\alpha^1_i\mp i\alpha^2_i\right]\ \ ,\ \ 
{\alpha^\pm_i}^\dagger=\frac{1}{\sqrt{2}}\left[{\alpha^1_i}^\dagger\pm 
i{\alpha^2_i}^\dagger\right]\ \ \ ,
\end{equation}
such that
\begin{equation}
\left[\alpha^+_i,{\alpha^+_j}^\dagger\right]=\delta_{ij}=
\left[\alpha^-_i,{\alpha^-_j}^\dagger\right]\ \ ,
\end{equation}
while one also has
\begin{equation}
z^\pm_i=\frac{1}{\sqrt{2}}\left[z^1_i\mp iz^2_i\right]\ \ ,\ \ 
\bar{z}^\pm_i=\frac{1}{\sqrt{2}}\left[\bar{z}^1_i\pm i\bar{z}^2_i\right]\ \ .
\end{equation}

Consequently, the following relations hold,
\begin{equation}
z^a_i{\alpha^a_i}^\dagger=z^+_i{\alpha^+_i}^\dagger+
z^-_i{\alpha^-_i}^\dagger\ \ ,\ \ 
|z|^2\equiv|z^a_i|^2=|z^+_i|^2+|z^-_i|^2\ \ ,
\end{equation}
as well as,
\begin{equation}
\hat{H}_0=\hbar g\omega\left[{\alpha^+_i}^\dagger\alpha^+_i+
{\alpha^-_i}^\dagger\alpha^-_i+d\right]=\hbar g\omega
\left[\hat{N}+d\right]\ \ ,\ \ 
\hat{\phi}=-\hbar\left[{\alpha^+_i}^\dagger\alpha^+_i
-{\alpha^-_i}^\dagger\alpha^-_i\right]\ \ .
\label{eq:H0helicity}
\end{equation}

Correspondingly, the helicity Fock state basis is given by
\begin{equation}
|n^\pm_i>=\prod_i\frac{1}{\sqrt{n^+_i!\ n^-_i!}}\,
\left({\alpha^+_i}^\dagger\right)^{n^+_i}
\left({\alpha^-_i}^\dagger\right)^{n^-_i}\,|0>\ \ \ ,
\end{equation}
$|0>$ being the usual normalised vacuum state, while the holomorphic 
helicity coherent states (strictly, holomorphic up to a factor) are defined by
\begin{equation}
|z^\pm_i>=e^{-\frac{1}{2}|z|^2}\,
e^{z^+_i{\alpha^+_i}^\dagger}\,
e^{z^-_i{\alpha^-_i}^\dagger}\,|0>\ \ \ .
\end{equation}

In particular, the Fock states $|n^\pm_i>$ diagonalise both the
Hamiltonian $\hat{H}_0$ and the gauge generator $\hat{\phi}$, since,
\begin{equation}
\hat{H}_0|n^\pm_i>=\hbar g\omega\left[\sum_i(n^+_i+n^-_i)+d\right]|n^\pm_i>
\ \ ,\ \ 
\hat{\phi}|n^\pm_i>=-\hbar\sum_i(n^+_i-n^-_i)\,|n^\pm_i>\ \ .
\end{equation}

Hence, the physical energy spectrum of gauge invariant states, 
namely those states 
annihilated by the operator $\hat{\phi}$, is given by
\begin{equation}
E_n=\hbar g\omega\,(2n+d)\ \ \ ,\ \ \ n=0,1,2,\dots\ \ \ ,
\label{eq:energy}
\end{equation}
and at energy level $E_n$, is spanned by those states $|n^\pm_i>$ 
which satisfy the left- and right-handed helicity matching condition,
\begin{equation}
\sum_i n^+_i=n=\sum_i n^-_i\ \ \ .
\label{eq:levelmatch}
\end{equation}
Nevertheless, this conclusion is a far cry from a complete characterization
of the physical spectrum of the model, since it does not provide
a rationale for the degeneracies in the energy spectrum when $d\ge 2$,
nor does it give any handle on the explicit contruction of these states.
As will be shown in the next three Sections, the physical projection
operator\cite{Klauder1} provides precisely the adequate tool to
address these issues, still entirely within Dirac's quantisation of the
model which does not entail any gauge fixing.

As a matter of fact, the advantages of the physical projector approach
are best exploited by working with the coherent state representation,
rather than the Fock space one. As a first illustration, let us consider
the action of the SO(2) gauge generator $\hat{\phi}$ on the helicity
coherent states $|z^\pm_i>$. In the same way as is demonstrated in Appendix A,
it is clear that for finite SO(2) transformations, one simply finds
\begin{equation}
e^{-\frac{i}{\hbar}\theta\hat{\phi}}\,|z^\pm_i>=
e^{i\theta[{\alpha^+_i}^\dagger\alpha^+_i
-{\alpha^-_i}^\dagger\alpha^-_i]}\,|z^\pm_i>=
|e^{\pm i\theta}z^\pm_i>\ \ \ ,
\end{equation}
or equivalently, in terms of the cartesian variables,
\begin{equation}
e^{-\frac{i}{\hbar}\theta\hat{\phi}}\,|z^a_i>=
|U^{ab}(\theta)z^b_i>\ \ \ ,
\end{equation}
$U^{ab}(\theta)$ being the SO(2) rotation matrix given in (\ref{eq:Uab}).
Since the physical projection operator involves precisely a gauge
invariant integration of all such finite transformations over the entire 
gauge group, these simple properties of coherent states clearly
indicate that they provide for an ideal calculational tool in this context.

\section{The SO(2) Model: the Gauge Invariant Partition Function}
\label{Sect5}

The physical projection operator\cite{Klauder1} is a quantum object 
which, being applied onto any quantity, constructs a gauge invariant one 
simply by averaging over the manifold of the gauge symmetry group
all finite gauge transformations generated by the first-class constraints 
of a system. Quite clearly, this averaging procedure projects out of 
any quantity its gauge variant components, while leaving unaffected 
its gauge invariant and thus physically observable ones. In particular,
acting on any given state in the representation space of the quantised
system, the physical projector constructs a gauge invariant state
annihilated by the gauge constraints, namely a physical state, hence its
name.

In the case of the SO(2) model, the group manifold is simply the unit
circle parametrised by the rotation angle $0\le\theta\le 2\pi$ introduced 
in (\ref{eq:Uab}) and associated to rotations in the index $a=1,2$.
Consequently, since the normalised invariant integration measure over the
SO(2) group reduces to $\int_0^{2\pi}d\theta/2\pi$, for this
model the physical projector is given by the following operator,
\begin{equation}
\proj=\frac{1}{2\pi}\int_0^{2\pi}\,d\theta\,
e^{-\frac{i}{\hbar}\theta\hat{\phi}}\ \ \ ,
\label{eq:proj}
\end{equation}
such that,
\begin{equation}
\proj\,^\dagger=\proj\ \ \ ,\ \ \ \proj\,^2=\proj\ \ \ .
\end{equation}
Obviously, the physical projector $\proj\,$ commutes with any gauge invariant
operator, such as the first-class Hamiltonian $\hat{H}_0$, and leaves
invariant any physical state annihilated by the gauge generator $\hat{\phi}$.

One direct application of the physical projector is in the description
of the time evolution of physical states. Given a specific
Lagrange multiplier function $\lambda(t)$, the time evolution operator
of the quantised system is given by\footnote{Since $\hat{H}_0$ and 
$\hat{\phi}$ have a vanishing commutator, the time-ordered exponential
indeed reduces to this expression.},
\begin{equation}
U(t_2,t_1)=e^{-\frac{i}{\hbar}\int_{t_1}^{t_2}dt
[\hat{H}_0+\lambda(t)\hat{\phi}]}\ \ \ .
\end{equation}
By definition, this evolution propagates in time gauge variant as well 
as gauge invariant states. It is in order to consistently propagate gauge
invariant states only that all other methods of quantisation have been
developed, whether the so-called reduced phase space ones or the
BFV-BRST extended ones\cite{Reviews}. With the help of the physical
projector however, a unitary consistent time propagation of physical
states only is readily achieved by projecting out any gauge variant
contribution, namely,
\begin{equation}
U_{\rm phys}(t_2,t_1)=U(t_2,t_1)\proj\,=\proj\ U(t_2,t_1)\proj\ \ \ .
\end{equation}
Moreover, due to the integration over the angle $\theta$ in (\ref{eq:proj}),
any contribution stemming from the Lagrange multiplier $\lambda(t)$
is averaged out as well, leaving the physical evolution operator built
from the first-class Hamiltonian $\hat{H}_0$ only, and thus finally
\begin{equation}
U_{\rm phys}(t_2,t_1)=e^{-\frac{i}{\hbar}(t_2-t_1)\hat{H}_0}\,\proj\,=\,
\proj\ e^{-\frac{i}{\hbar}(t_2-t_1)\hat{H}_0}\,\proj\ \ \ .
\end{equation}

This form makes it clear that the spectral decomposition of the physical
evolution operator $U_{\rm phys}(t_2,t_1)$ reads as follows,
\begin{equation}
U_{\rm phys}(t_2,t_1)=\sum_{E_n,\mu_n}\,
e^{-\frac{i}{\hbar}(t_2-t_1)E_n}\,|E_n,\mu_n><E_n,\mu_n|\ \ \ ,
\end{equation}
where $|E_n,\mu_n>$ are the orthonormalised physical states of energy
$E_n=\hbar g\omega(2n+d)$, $\mu_n$ being a multi-index labelling the
associated energy degeneracies. In view of the energy spectrum of physical 
states, this same expression also reads as
\begin{equation}
U_{\rm phys}(t_2,t_1)=e^{-i(t_2-t_1)g\omega d}\,
\sum_{n,\mu_n}\,e^{-i(t_2-t_1)g\omega(2n)}\,|E_n,\mu_n><E_n,\mu_n|\ \ \ .
\end{equation}

Consequently, given an arbitrary complex parameter\footnote{Possibly 
such that $|x|<1$ in order to ensure uniform convergence of the quantity 
to be introduced presently.} $x$, the basic operator which encompasses all the
properties of gauge invariant physical states is\footnote{While the
physical evolution operator $U_{\rm phys}(t_2,t_1)$
thus corresponds to $x^{\hat{N}+d}\proj\,$ evaluated for 
$x=e^{-i(t_2-t_1)g\omega}$.}
\begin{equation}
\sum_{n,\mu_n}\,x^{2n}\,|E_n,\mu_n><E_n,\mu_n|\,=
\,\proj\ x^{\hat{N}}\proj\,=\,x^{\hat{N}}\proj\ \ \ ,
\label{eq:partf}
\end{equation}
$\hat{N}$ being the total excitation level operator introduced
in (\ref{eq:Nhat}) and (\ref{eq:H0helicity}). 
Indeed, the trace of that operator is the partition
function for physical states, since, based on the above identification and
for $|x|<1$, we have
\begin{equation}
{\rm Tr}\,x^{\hat{N}}\proj=\sum_{n=0}^\infty\,d_n\,x^{2n}\ \ \ ,
\end{equation}
where the coefficients $d_n$ $(n=0,1,2\dots)$ are the degeneracies of
physical states at excitation level $\hat{N}=2n$, thus at the energy level 
$E_n=\hbar g\omega(2n+d)$.

Similarly, even only diagonal matrix elements of the operator
(\ref{eq:partf}) provide a generating function for the wave functions
of the gauge invariant states. For example in the helicity coherent
state basis, one has
\begin{equation}
<z^\pm_i|x^{\hat{N}}\proj\,|z^\pm_i>=
\sum_{n,\mu_n}\,x^{2n}\,|<z^\pm_i|E_n,\mu_n>|^2\ \ \ .
\label{eq:generatingwave}
\end{equation}
Hence, up to a constant phase, which is always a matter of convention
anyway, wave functions of physical states may be determined from
the diagonal matrix elements of the operator (\ref{eq:partf}).

In conclusion, all that remains to be done in order to solve the physical 
spectrum of the system and its time evolution, is the evaluation of the
operator (\ref{eq:partf}), for which the coherent state representation
is most convenient for reasons already pointed out in the previous
Section.

In particular, let us consider the determination of the partition function.
Since
\begin{equation}
{\rm Tr}\,x^{\hat{N}}\proj\,=\int_0^{2\pi}\frac{d\theta}{2\pi}
\int\prod_{\pm,i}\frac{dz^\pm_id\bar{z}^\pm_i}{\pi}
<z^\pm_i|x^{\hat{N}}e^{i\theta[{\alpha^+_i}^\dagger\alpha^+_i-
{\alpha^-_i}^\dagger\alpha^-_i]}|z^\pm_i>\ \ \ ,
\end{equation}
one simply has
\begin{equation}
{\rm Tr}\,x^{\hat{N}}\proj\,=\int_0^{2\pi}\frac{d\theta}{2\pi}
\int\prod_{\pm,i}\frac{dz^\pm_id\bar{z}^\pm_i}{\pi}
<z^\pm_i|xe^{\pm i\theta}z^\pm_i>\,e^{-\frac{1}{2}(1-|x|^2)|z|^2}\ \ \ .
\end{equation}
The gaussian integrals stemming from the coherent state matrix elements
are readily performed (see Appendix A), leading to
\begin{equation}
{\rm Tr}\,x^{\hat{N}}\proj\,=\int_0^{2\pi}\frac{d\theta}{2\pi}\,
\frac{1}{\left[1-xe^{i\theta}\right]^d\,\left[1-xe^{-i\theta}\right]^d}\ \ \ ,
\label{eq:partfdeterminant}
\end{equation}
and thus, finally,
\begin{equation}
{\rm Tr}\,x^{\hat{N}}\proj\,=\sum_{n=0}^\infty\,d_n\,x^{2n}\ \ \ ,
\end{equation}
with the degeneracies given by
\begin{equation}
d_n=\left[\frac{(d-1+n)!}{(d-1)!\,n!}\right]^2\ \ \ ,\ \ \ n=0,1,2,\dots\ \ \ .
\label{eq:degeneracies}
\end{equation}

In the case $d=1$, these degeneracies $d_n=1$ are trivial at all physical
excitations levels $n=0,1,2,\dots$, a conclusion which is indeed obvious
given the left- and right-helicity matching condition (\ref{eq:levelmatch}) 
on the physical spectrum.

When $d=2$, the degeneracies $d_n=(n+1)^2$ as well as the energy
spectrum $E_n=2\hbar g\omega(n+1)$, do indeed agree with those 
determined in Ref.\cite{Pause}. However, note how by having introduced
the physical projector $\proj\,$, all degeneracies $d_n$ are readily 
obtained whatever the number $d$ of particles involved in two dimensions, 
without the necessity of considering the subtle issues of
gauge fi\-xing, modular space and the spectrum of the ensuing
Schr\"odinger equation, aspects which are all unavoidable in 
the standard approach adopted in Ref.\cite{Pause}.

\section{The SO(2) Model: the SO($d$) Valued Partition Function}
\label{Sect6}

The ever increasing degeneracies (\ref{eq:degeneracies}) with energy
level beg for an understanding. A remark which goes part way towards
a complete explanation is to notice that all $d$, two dimensional
spherical harmonic oscillators share a common angular frequency $\omega$.
In other words, in addition to the SO(2) local gauge symmetry
associated to the index $a=1,2$, the system also possesses a global
SO($d$) symmetry associated to rotations in the index $i=1,2,\cdots,d$.
However, as we shall see, this property still does not account completely
for all the observed degeneracies.

Therefore, all gauge invariant physical states must also fall into different
representations of the global SO($d$) symmetry, thereby explaining
(to some extent) the obtained degeneracies. It should thus be possible
to ``tag" each of the physical states in terms both of the SO($d$) 
representation to which it belongs and its quantum numbers under that 
symmetry.  This is possible by extending the operator $x^{\hat{N}}\proj\,$ 
to include the SO($d$) transformation properties of physical states, 
namely by considering the SO($d$) generators of the system.

Clearly, in terms of the cartesian or helicity quantum degrees of freedom,
these $d(d-1)/2$ SO($d$) generators are given by
\begin{equation}
\hat{L}_{ij}=i\hbar\left[{\alpha^a_i}^\dagger\alpha^a_j-
{\alpha^a_j}^\dagger\alpha^a_i\right]=
i\hbar\left[{\alpha^+_i}^\dagger\alpha^+_j+
{\alpha^-_i}^\dagger\alpha^-_j-
{\alpha^+_j}^\dagger\alpha^+_i-
{\alpha^-_j}^\dagger\alpha^-_i\right]\ \ \ ,
\end{equation}
and thus satisfy the algebra
\begin{equation}
\left[\hat{L}_{ij},\hat{L}_{kl}\right]=-i\hbar\left[
\delta_{ik}\hat{L}_{jl}-
\delta_{il}\hat{L}_{jk}-
\delta_{jk}\hat{L}_{il}+
\delta_{jl}\hat{L}_{ik}\right]\ \ \ .
\end{equation}

Related to these expressions, the matrix representations of the
SO($d$) generators in the de\-fi\-ning $d$-dimensional representation are
given by
\begin{equation}
\left(T_{ij}\right)_{kl}=i\hbar\left[\delta_{ik}\delta_{jl}-
\delta_{il}\delta_{jk}\right]\ \ \ ,
\end{equation}
so that in fact $\hat{L}_{ij}=\alpha^\dagger\cdot T_{ij}\cdot\alpha$,
with the contraction operating on the indices $k,l=1,2,\cdots,d$ and 
a diagonal summation over the SO(2) gauge index $a=1,2$.
Finite SO($d$) transformations in the $d$-dimensional vector
representation are then given by the $d\times d$ rotation matrices,
\begin{equation}
R_{kl}\left(\omega_{ij}\right)=
\left(e^{-\frac{i}{2\hbar}\omega_{ij}T_{ij}}\right)_{kl}\ \ \ ,
\end{equation}
where $\omega_{ij}$ are $d(d-1)/2$ angular variables parametrising
the SO($d)$ group manifold.

Consequently (see Appendix A), creation operators transform as follows
under the global SO($d$) symmetry of the model,
\begin{equation}
e^{-\frac{i}{2\hbar}\omega_{ij}\hat{L}_{ij}}\,{\alpha^\pm_i}^\dagger\,
e^{\frac{i}{2\hbar}\omega_{ij}\hat{L}_{ij}}=
{\alpha^\pm_j}^\dagger\,R_{ji}\left(\omega_{ij}\right)\ \ \ ,
\end{equation}
while for coherent states, one has
\begin{equation}
e^{-\frac{i}{2\hbar}\omega_{ij}\hat{L}_{ij}}\,|z^\pm_i>=
|R_{ij}\left(\omega_{ij}\right)\,z^\pm_j>\ \ \ .
\end{equation}

Given these considerations, the generalisation of the operator 
$x^{\hat{N}}\proj\,$ to an SO($d$) valued ope\-ra\-tor, which ``tags" physical 
states by their SO($d$) representation quantum numbers, is simply defined by
\begin{equation}
x^{\hat{N}}e^{-\frac{i}{2\hbar}\omega_{ij}\hat{L}_{ij}}\,\proj\,\ \ \ .
\label{eq:partfSO(d)}
\end{equation}
For the same reasons as detailed in the previous Section, the trace
of this operator provides the SO($d$) valued partition function of
physical states, while its diagonal matrix elements generate their
SO($d$) valued wave functions. Consequently,
the operator (\ref{eq:partfSO(d)}) is the ideal tool for determining
the SO($d$) characterization of gauge invariant physical states.

Considering then the SO($d$) valued partition function, one simply finds
again,
\begin{equation}
{\rm Tr}\,x^{\hat{N}}e^{-\frac{i}{2\hbar}\omega_{ij}\hat{L}_{ij}}\proj\,=
\int_0^{2\pi}\frac{d\theta}{2\pi}\int\prod_{\pm,i}
\frac{dz^\pm_id\bar{z}^\pm_i}{\pi}\,
<z^\pm_i|xe^{\pm i\theta}R_{ij}\left(\omega_{ij}\right)\,z^\pm_j>\,
e^{-\frac{1}{2}(1-|x|)^2)|z|^2}\ \ \ ,
\end{equation}
so that finally,
\begin{equation}
{\rm Tr}\,x^{\hat{N}}e^{-\frac{i}{2\hbar}\omega_{ij}\hat{L}_{ij}}\proj\,=
\int_0^{2\pi}\frac{d\theta}{2\pi}\,\frac{1}
{{\rm det}\left[\delta_{ij}-xe^{i\theta}R_{ij}\left(\omega_{ij}\right)\right]\,
{\rm det}\left[\delta_{ij}-xe^{-i\theta}R_{ij}\left(\omega_{ij}\right)\right]}
\ \ \ ,
\label{eq:partfSO(d)det}
\end{equation}
a result which generalises in a transparent way the one obtained
previously for the partition function in (\ref{eq:partfdeterminant}).

The evaluation of this expression for an arbitrary choice of angular
variables $\omega_{ij}$ would be quite involved. However, given our
purpose of determining the SO($d$) representations of physical states,
it suffices to only consider a maximal commuting subalgebra among the
generators $\hat{L}_{ij}$, namely the
Cartan subalgebra\cite{Slansky}. Indeed, representations of compact
semi-simple Lie algebras may be characterized by the Dynkin labels of the
Dynkin diagram related to the Cartan subalgebra. In the
case of the SO($d$) group, two general classes have to be distinguished
according to whether $d$ is even or odd. Consequently,
in order to proceed with the calculation of the SO($d$) valued partition 
function restricted to the Cartan subalgebra, it proves useful to first 
consider the simple cases with $d=1$ and $d=2$, which will display
the structure of the general solution.

The case $d=1$ is trivial, since no global symmetry is then present,
and the SO($d=1$) valued partition function thus reduces to the
simple partition function (\ref{eq:partfdeterminant}), namely,
\begin{equation}
d=1:\ \ \ \ \ \ \ 
{\rm Tr}\,x^{\hat{N}}\proj\,=
\int_0^{2\pi}\frac{d\theta}{2\pi}\,\frac{1}
{\left[1-xe^{i\theta}\right]\left[1-xe^{-i\theta}\right]}=
\sum_{n=0}^\infty x^{2n}=\frac{1}{1-x^2}\ \ \ .
\end{equation}

Consider now the $d=2$ case. Here again, the appearance of the global
SO(2)=U(1) symmetry in the index $i=1,2$ suggests the introduction of
helicity-like degrees of freedom, such as ${\alpha^\pm_\pm}^\dagger$ and
$\alpha^\pm_\pm$, as well as $z^\pm_\pm$, defined in terms of complex linear 
combinations of the cartesian creation and annihilation operators 
${\alpha^\pm_i}^\dagger$ and $\alpha^\pm_i$ ($i=1,2$) (in the same way
as were the latter helicity operators defined in terms of 
${\alpha^a_i}^\dagger$ and $\alpha^a_i$ $(a=1,2)$), 
with thus the associated helicity coherent 
states $|z^\pm_\pm>$ (see also Appendix A). Since the procedure should 
by now be clear, it is not spelled out explicitly here.

On the other hand, the SO(2) algebra is abelian with only one generator
$\hat{L}_{12}$, thus defining in a trivial manner the Cartan subalgebra.
This operator being given by (a summation over the upper index $\pm$ is
implicit)
\begin{equation}
\hat{L}_{12}=i\hbar\left[{\alpha^\pm_1}^\dagger\alpha^\pm_2-
{\alpha^\pm_2}^\dagger\alpha^\pm_1\right]=
-\hbar\left[{\alpha^\pm_+}^\dagger\alpha^\pm_+-
{\alpha^\pm_-}^\dagger\alpha^\pm_-\right]\ \ \ ,
\end{equation}
its action on the coherent states $|z^\pm_\pm>$ is obviously
\begin{equation}
e^{-\frac{i}{2\hbar}\omega_{ij}\hat{L}_{ij}}\,|z^\pm_\pm>=
|e^{\pm i\omega_{12}}z^\pm_\pm>\ \ \ ,
\end{equation}
where the $\pm$ sign in the phase factor on the r.h.s. is of course
correlated with that of the {\em lower\/} index of $z^\pm_\pm$.

Consequently, it should be clear that for $d=2$, by working in the $i=1,2$
helicity basis, the SO($d=2$) valued partition function 
(\ref{eq:partfSO(d)det}) reduces to,
\begin{equation}
{\rm Tr}\,x^{\hat{N}}e^{-\frac{i}{2\hbar}\omega_{ij}\hat{L}_{ij}}\proj\,=
\int_0^{2\pi}\frac{d\theta}{2\pi}\,\frac{1}
{\left[1-xe^{i(\theta+\omega_{12})}\right]
\left[1-xe^{i(\theta-\omega_{12})}\right]
\left[1-xe^{-i(\theta-\omega_{12})}\right]
\left[1-xe^{-i(\theta+\omega_{12})}\right]}\ \ \ .
\end{equation}
Through a careful analysis of the
series expansion of this integral in powers of $x$, one finally obtains
\begin{equation}
d=2:\ \ \ \ \
{\rm Tr}\,x^{\hat{N}}e^{-\frac{i}{2\hbar}\omega_{ij}\hat{L}_{ij}}\proj\,=
\sum_{n=0}^\infty\,x^{2n}\,\sum_{p=-n}^n\,\Big[(n+1)-|p|\Big]\,
e^{2ip\omega_{12}}\ \ \ .
\label{eq:partitiond2}
\end{equation}
This result thus establishes that for $d=2$,
all $d_n=(n+1)^2$ physical states at energy level 
$E_n=2\hbar g\omega(n+1)$ $(n=0,1,\dots)$ fall into the one dimensional
representations of the SO($d=2$)=U(1) global symmetry
of integer helicity $-n\le p\le n$, with a degeneracy $d(n,p)=(n+1-|p|)$
for each of these helicity representations in the index $i=1,2$.
In particular, we have indeed that,
\begin{equation}
d=2:\ \ \ \ \ 
\sum_{p=-n}^n\,d(n,p)=(n+1)^2=d_n\ \ \ ,\ \ \ n=0,1,2,\dots\ \ \ .
\end{equation}
In conclusion, the existence of the global SO($d=2$) symmetry does account
for the degeneracies in the energy spectrum of physical states,
though not completely, since the SO($d=2$) representations themselves 
possess some degree of degeneracy which still needs to be understood. 
As a matter of fact, remarks to that effect are made in the Conclusions, 
without attempting a general solution to this issue in the present paper.

Having solved the $d=2$ case, let us now turn first to the general situation
when $d=2M$ is even. The Cartan subalgebra of the SO($d=2M$) global
symmetry may then be described as follows. 
The $2M$ indices $i=1,2,\cdots,d(=2M)$
may be grouped into $M$ consecutive pairs $(1,2)$, $(3,4)$, $\dots$,
$(2M-1,2M)$, to be labelled by an index $\mu=1,2,\cdots,M$ in the same
order (thus the index $\mu$ refers to the pair
$(i=2\mu-1,j=2\mu)$). Correspondingly, the Cartan subalgebra is 
spanned by the SO($d=2M$) generators $\hat{L}_{ij}$ associated
to this organisation in the $(i,j)$ indices, with the identification,
\begin{equation}
\hat{L}_\mu=\hat{L}_{2\mu-1,2\mu}\ \ \ ,\ \ \ \mu=1,2,\dots,M\ \ \ ,
\end{equation}
thereby leading to $M$ commuting generators in SO($d=2M$), which is
indeed the rank of the algebra of that group. In particular, finite group
transformations restricted to the corresponding maximal torus in
SO($d=2M$) are generated by,
\begin{equation}
e^{-\frac{i}{2\hbar}\omega_{ij}\hat{L}_{ij}}=
e^{-\frac{i}{\hbar}\omega_\mu\hat{L}_\mu}\ \ \ ,
\end{equation}
with the following association in the angular parameters $\omega_\mu$
and $\omega_{ij}$,
\begin{equation}
\omega_\mu=\omega_{2\mu-1,2\mu}\ \ \ ,\ \ \ \mu=1,2,\cdots,M\ \ \ .
\end{equation}

Clearly, through this construction we have achieved for each value of
$\mu=1,2,\cdots,M$, a situation identical to that developed previously
for $d=2$, to which the helicity techniques may again be applied. 
Hence, the Cartan subalgebra restricted SO($d=2M$) valued
partition function (\ref{eq:partfSO(d)det}) simply reads,
\begin{equation}
{\rm Tr}\,x^{\hat{N}}e^{-\frac{i}{\hbar}\omega_\mu\hat{L}_\mu}\proj\,=
\int_0^{2\pi}\frac{d\theta}{2\pi}\,\prod_{\mu=1}^M\,\frac{1}
{\left[1-xe^{i(\theta+\omega_\mu)}\right]
\left[1-xe^{i(\theta-\omega_\mu)}\right]
\left[1-xe^{-i(\theta-\omega_\mu)}\right]
\left[1-xe^{-i(\theta+\omega_\mu)}\right]}\ \ \ ,
\end{equation}
reducing finally to the expression,
\begin{equation}
\sum_{n^+_\mu,n^-_\mu,m^+_\mu,m^-_\mu=0}^\infty
x^{\sum_{\mu=1}^M(n^+_\mu+n^-_\mu+m^+_\mu+m^-_\mu)}\,
e^{i\sum_{\mu=1}^M\omega_\mu(n^+_\mu-n^-_\mu-m^+_\mu+m^-_\mu)}\,
\delta\left(\sum_{\mu=1}^M(n^+_\mu-n^-_\mu+m^+_\mu-m^-_\mu)\right)\ \ \ .
\label{eq:partf2M}
\end{equation}
In this power series expansion in $x$, the power of $x$ determines
the excitation level of a contributing gauge invariant physical state,
while the Kronecker $\delta$ constraint corresponds to the left- and
right-handed SO(2) gauge helicity matching condition (\ref{eq:levelmatch})
for that state. Finally, the SO($d=2M$) properties of physical states
are encoded in the phase factor involving the parameters $\omega_\mu$.
The coefficients of these parameters determine the SO($d=2M$)
quantum numbers of these states, {\em i.e.\/} the Dynkin labels
of the weight vectors associated to each one of these states as members of 
specific SO($d=2M$) representations. In particular, the phase factors for
highest weight states determine the Dynkin labels of the SO($d=2M$)
representations contributing to the partition function. In this manner,
starting with the very highest weight state and identifying its SO($d=2M$)
descendants, and so on by recursion starting again from the remaining
highest weight states, in principle it is possible to identify the
content of SO($d=2M$) representations appearing at each energy level
in the physical spectrum, including their degeneracies. 
However, based on the $d=2$ explicit solution,
one expects that this classification does not exhaust completely the
degeneracies of the energy spectrum, for which a deeper understanding
is still called for in terms of a symmetry larger than SO($d=2M$). 
Since such an analysis is deferred to later work, the complete 
characterization of the SO($d=2M$) representations for physical states
enscribed in (\ref{eq:partf2M}) is not pursued further here.

Finally, turning to the case when $d=2M+1$ is odd, it should now be
clear how to proceed. Simply, all $2M+1$ indices $i=1,2,\cdots,(d=2M+1)$
are again grouped into consecutive pairs $(1,2)$, $(3,4)$, $\cdots$,
$(2M-1,2M)$, labelled by $\mu=1,2,\cdots,M$ in the same
order and thus leaving the odd index $i=2M+1$ on its own.
Correspondingly, the Cartan subalgebra is then obtained through the
identification
\begin{equation}
\hat{L}_\mu=\hat{L}_{2\mu-1,2\mu}\ \ \ ,\ \ \ \mu=1,2,\cdots,M\ \ \ ,
\end{equation}
hence $M$ such commuting SO($d=2M+1$) generators, which is indeed
the rank of the algebra of the group SO($d=2M+1$). Clearly here again,
each of the $M$ pairs $(i=2\mu-1,j=2\mu)$ is identical to the
$d=2$ system, while the remaining index $i=2M+1$ may be viewed as
the $d=1$ system.

As a result, given the identification of SO($d=2M+1$) angular parameters
$\omega_\mu=\omega_{2\mu-1,2\mu}$, the Cartan subalgebra restricted
SO($d=2M+1$) valued partition function simply reads
\begin{displaymath}
{\rm Tr}\,x^{\hat{N}}e^{-\frac{i}{\hbar}\omega_\mu\hat{L}_\mu}\proj\,=
\end{displaymath}
\begin{equation}
\int_0^{2\pi}\frac{d\theta}{2\pi}\frac{1}
{\left[1-xe^{i\theta}\right]\left[1-xe^{-i\theta}\right]}\,\prod_{\mu=1}^M
\frac{1}
{\left[1-xe^{i(\theta+\omega_\mu)}\right]
\left[1-xe^{i(\theta-\omega_\mu)}\right]
\left[1-xe^{-i(\theta-\omega_\mu)}\right]
\left[1-xe^{-i(\theta+\omega_\mu)}\right]}\ \ \ .
\end{equation}
Again, this result may be expressed through a power series expansion
in $x$, with phase factors determining the SO($d=2M+1$)
quantum numbers of gauge invariant physical states. However, for the
same reasons as those given for the $d=2M$ case, this point shall not be
pursued any further in this paper.

Nevertheless, the discussion of this Section has demonstrated unequivocally,
that the physical projector approach allows for a complete characterization
of all the symmetry properties of the physical spectrum of a gauge
invariant theory. The next Section addresses the determination of the
associated wave functions through the same calculational tools.

\section{The SO(2) Model: Wave Functions of Physical States}
\label{Sect7}

As pointed out in (\ref{eq:generatingwave}), the determination of
physical state wave functions entails simply the eva\-lu\-a\-tion of the
diagonal matrix elements of the operator $x^{\hat{N}}\proj\,$,
which are most straightforwardly obtained in the coherent state basis.
Based on the considerations of the previous Section, this
ge\-ne\-rating function for physical state wave functions may be extended
by introducing a ``tagging" in terms of representations of the SO($d$)
global symmetry, for which it suffices to restrict SO($d$) group
elements to the maximal torus defined by the Cartan subalgebra.
In so doing, wave functions will appear accompanied in the generating
function by the relevant phase factors which characterize their
SO($d$) quantum numbers, thereby enabling a direct identification
of the appropriate wave functions which otherwise would appear only
through linear combinations due to the energy degeneracies. Nevertheless,
degeneracies in the SO($d$) representations themselves could still
lead to ambiguities in this identification of wave functions, if
only in their proper normalisation\footnote{Indeed, the generating function
should provide not only the wave functions, but also their proper
normalisation.}, even though the determination of these degeneracies
from the partition function along the lines outlined in the previous
Section should allow to lift such ambiguities. Clearly, once identified,
the symmetry beyond SO($d$) accounting completely for all such degeneracies
would circumvent these issues altogether.

For the time being though, let us consider the following coherent state
matrix elements,
\begin{equation}
<z^\pm_i|x^{\hat{N}}
e^{-\frac{i}{2\hbar}\omega_{ij}\hat{L}_{ij}}\,\proj\,|z^\pm_i>\ \ \ ,
\end{equation}
where the finite SO($d$) group elements are restricted to the maximal torus
defined by Cartan subalgebra. Given the developments of the previous
Sections, it is clear that these matrix elements read as
\begin{equation}
<z^\pm_i|x^{\hat{N}}
e^{-\frac{i}{2\hbar}\omega_{ij}\hat{L}_{ij}}\,\proj\,|z^\pm_i>=
\int_0^{2\pi}\frac{d\theta}{2\pi}\,e^{-|z|^2}\,\prod_{\pm}\,
e^{xe^{\pm i\theta}\bar{z}^\pm_i R_{ij}(\omega_{ij})z^\pm_j}\ \ \ .
\label{eq:wavefSO(d)}
\end{equation}

Let us first assume $d=2M$ to be even. Given the
Cartan subalgebra of SO($d=2M$) described in the previous Section, 
and the choice of the corresponding helicity coordinate basis in the
index $i=1,2,\cdots,(d=2M)$, (\ref{eq:wavefSO(d)}) becomes,
\begin{equation}
\int_0^{2\pi}\frac{d\theta}{2\pi}e^{-|z|^2}\,\prod_{\pm}\left\{
\prod_{\mu=1}^M
e^{xe^{\pm i\theta}\left[e^{i\omega_\mu}|z^\pm_{\mu+}|^2+
e^{-i\omega_\mu}|z^\pm_{\mu-}|^2\right]}\right\}\ \ \ ,
\end{equation}
where,
\begin{equation}
z^\pm_{\mu+}=\frac{1}{\sqrt{2}}\left[z^\pm_{2\mu-1}-iz^\pm_{2\mu}\right]
\ \ \,\ \ \ 
z^\pm_{\mu-}=\frac{1}{\sqrt{2}}\left[z^\pm_{2\mu-1}+iz^\pm_{2\mu}\right]
\ \ \ .
\end{equation}
Upon expansion in $x$, one thus obtains for the Cartan subalgebra restricted
SO($d=2M$) valued generating function of physical state wave functions,
\begin{equation}
\sum_{n=0}^\infty x^{2n}\,\frac{e^{-|z|^2}}{(n!)^2}\,
\left\{\sum_{\mu,\nu=1}^M
\left(e^{i\omega_\mu}|z^+_{\mu+}|^2+e^{-i\omega_\mu}|z^+_{\mu-}|^2\right)
\left(e^{i\omega_\nu}|z^-_{\nu+}|^2+e^{-i\omega_\nu}|z^-_{\nu-}|^2\right)
\right\}^n\ \ \ .
\end{equation}
Note that in this last expression, the term in curly brackets (raised to 
the $n$-th power) may  be expressed as 
\begin{equation}
\sum_{\mu,\nu=1}^M\left(
e^{i(\omega_\mu+\omega_\nu)}|z^+_{\mu+}z^-_{\nu+}|^2+
e^{i(\omega_\mu-\omega_\nu)}|z^+_{\mu+}z^-_{\nu-}|^2+
e^{-i(\omega_\mu-\omega_\nu)}|z^+_{\mu-}z^-_{\nu+}|^2+
e^{-i(\omega_\mu+\omega_\nu)}|z^+_{\mu-}z^-_{\nu-}|^2\right)\ \ \ .
\label{eq:expansionwave2M}
\end{equation}

Similarly in the case $d=2M+1$, (\ref{eq:wavefSO(d)}) is given by,
\begin{equation}
\int_0^{2\pi}\frac{d\theta}{2\pi}e^{-|z|^2}\,\prod_\pm\left\{
e^{xe^{\pm i\theta}|z^\pm_{2M+1}|^2}\,
\prod_{\mu=1}^M e^{xe^{\pm i\theta}\left[e^{i\omega_\mu}|z^\pm_{\mu+}|^2+
e^{-i\omega_\mu}|z^\pm_{\mu-}|^2\right]}\right\}\ \ \ ,
\end{equation}
which leads to the series representation,
\begin{displaymath}
\sum_{n=0}^\infty x^{2n}\,\frac{e^{-|z|^2}}{(n!)^2}\times
\end{displaymath}
\begin{equation}
\times\left\{\left(
\sum_{\mu=1}^M(e^{i\omega_\mu}|z^+_{\mu+}|^2+e^{-i\omega_\mu}|z^+_{\mu-}|^2)
+|z^+_{2M+1}|^2\right)\left(
\sum_{\mu=1}^N(e^{i\omega_\nu}|z^-_{\nu+}|^2+e^{-i\omega_\nu}|z^-_{\nu-}|^2)
+|z^-_{2M+1}|^2\right)
\right\}^n\ \ \ .
\end{equation}
As in (\ref{eq:expansionwave2M}), the term in curly 
brackets may be expanded as follows,
\begin{displaymath}
\sum_{\mu,\nu=1}^M\left(
e^{i(\omega_\mu+\omega_\nu)}|z^+_{\mu+}z^-_{\nu+}|^2+
e^{i(\omega_\mu-\omega_\nu)}|z^+_{\mu+}z^-_{\nu-}|^2+
e^{-i(\omega_\mu-\omega_\nu)}|z^+_{\mu-}z^-_{\nu+}|^2+
e^{-i(\omega_\mu+\omega_\nu)}|z^+_{\mu-}z^-_{\nu-}|^2\right)\,+
\end{displaymath}
\begin{displaymath}
+\,\sum_{\mu=1}^M\left(
e^{i\omega_\mu}|z^+_{\mu+}z^-_{2M+1}|^2+
e^{-i\omega_\mu}|z^+_{\mu-}z^-_{2M+1}|^2+
e^{i\omega_\mu}|z^+_{2M+1}z^-_{\mu+}|^2+
e^{-i\omega_\mu}|z^+_{2M+1}z^-_{\mu-}|^2\right)\,+\,
\end{displaymath}
\begin{equation}
+\,|z^+_{2M+1}z^-_{2M+1}|^2\ \ \ .
\label{eq:expansionwave2M+1}
\end{equation}

Recalling the arguments leading to (\ref{eq:generatingwave}), the
modulus squared wave functions of gauge in\-va\-riant physical states
at energy level $E_n$ are thus obtained by expanding the $n$-th
power of the sums (\ref{eq:expansionwave2M}) and
(\ref{eq:expansionwave2M+1}). It is clear that these expansions
lead indeed to sums of modulus squared products of polynomials
in the coherent state labels $z^\pm_i$, thereby also enabling
a direct identification of the combination of Fock
state basis vectors associated to these physical states, 
including their proper normalisation. The appropriate
linear combinations of these terms which are to be associated to given 
physical states are (partly) determined on the basis of their SO($d$) quantum
numbers ``tagged" by the Dynkin label phase factors $e^{\pm i\omega_\mu}$.
In the present analysis, only the remaining degeneracies in the SO($d$) 
representations themselves---rather than their state content---could lead 
to some ambiguities in these identifications, and if these degeneracies are
known on basis of the SO($d$) valued partition function, the ambiguities
would only be in the normalisation of the states which then has to be
calculated separately. However, as was pointed out previously, the 
identification of the complete symmetry accounting for all observed 
degeneracies should lift any such remaining ambiguity, including the
proper normalisation of all physical states.

The general structure of the expressions (\ref{eq:expansionwave2M}) and
(\ref{eq:expansionwave2M+1}) is also quite interesting. Indeed, each
modulus squared term involves the product of two $z^\pm_i$ factors
whose upper indices in $a=1,2$ are opposite in the helicity basis.
This fact makes the SO(2) gauge invariance of the associated states
indeed obvious, since the corresponding SO(2) charge is then identically
vanishing. Similarly, the structure in the lower index $i=1,2,\cdots,d$
of the products of the $z^\pm_i$ factors determines directly the corresponding
SO($d$) symmetry properties under transformations in the maximal torus 
generated by the Cartan subalgebra. Hence, upon expansion of the $n$-th 
power of (\ref{eq:expansionwave2M}) and (\ref{eq:expansionwave2M+1}) leading 
to specific binomial coefficients directly related to the normalisation
of their wave functions, the SO($d$) symmetry properties of physical
states may be directly read off the combination of lower indices
appearing in a given product of the $z^\pm_i$ factors.

It is useful to consider the cases $d=1$ and $d=2$ in some detail.
When $d=1$, the generating function of physical state wave functions
is simply
\begin{equation}
<z^\pm|x^{\hat{N}}\proj\,|z^\pm>=\sum_{n=0}^\infty\,
\frac{x^{2n}}{\left(n!\right)^2}
\,|z^+z^-|^{2n}\,e^{-|z^+|^2-|z^-|^2}\ \ \ .
\end{equation}
Consequently, the helicity coherent state wave functions of the single
physical state at each energy level $E_n=\hbar g\omega(2n+1)$ are given by
\begin{equation}
<E_n|z^\pm>=\frac{1}{n!}\,\left(z^+z^-\right)^n\,
e^{-\frac{1}{2}\left[|z^+|^2+|z^-|^2\right]}\ \ \ ,
\end{equation}
thus corresponding to the following orthonormalised Fock states
\begin{equation}
|E_n>=\frac{1}{n!}\left({\alpha^+}^\dagger\right)^n\,
\left({\alpha^-}^\dagger\right)^n\,|0>\ \ \ .
\end{equation}
Of course, in the case $d=1$, these results could have been obtained
directly from the energy spectrum (\ref{eq:energy}) and the level
matching condition (\ref{eq:levelmatch}).

In the $d=2$ case, the physical wave function generating function is
given by\footnote{The single value of the index $\mu=1$ is this case
is not displayed, while $|z|^2$ stands for the sum
$\sum_{\pm}(|z^\pm_+|^2+|z^\pm_-|^2)$.}
\begin{displaymath}
<z^\pm_\pm|x^{\hat{N}}e^{-\frac{1}{2}\omega_{ij}\hat{L}_{ij}}\proj\,
|z^\pm_\pm>=\sum_{n=0}^\infty\,\frac{x^{2n}}{\left(n!\right)^2}\,
e^{-|z|^2}\times
\end{displaymath}
\begin{equation}
\times\left\{
e^{2i\omega_{12}}|z^+_+z^-_+|^2\,+\,
\left|\frac{1}{\sqrt{2}}\left(z^+_+z^-_- + z^+_-z^-_+\right)\right|^2\,+\,
\left|\frac{1}{\sqrt{2}}\left(z^+_+z^-_- - z^+_-z^-_+\right)\right|^2\,+\,
e^{-2i\omega_{12}}|z^+_-z^-_-|^2\right\}^n\ \ \ ,
\label{eq:wavefSO(2)}
\end{equation}
where the two middle terms in (\ref{eq:expansionwave2M}) with vanishing
Dynkin label have been expressed as symmetric and antisymmetric 
combinations under the exchange of the two harmonic
oscillators $i=1$ and $i=2$. In this way, the identification of physical
states and of their coherent state wave functions is straightforward enough, 
without the possibility of any ambiguity\footnote{The real reason
for this fact will be made clear in the Conclusions.}. Thus, for example,
at the first two energy levels $n=0$ and $n=1$, the properly normalised
physical wave functions are simply
\begin{equation}
\begin{array}{r c l}
E_0=2\hbar g\omega\ &:&\ \ \ 
<E_0,p=0|z^\pm_\pm>=
e^{-\frac{1}{2}\left[|z^+_+|^2+|z^+_-|^2+|z^-_+|^2+|z^-_-|^2\right]}\ \ \ ,\\ \\
E_1=4\hbar g\omega\ &:&\ \ \ 
<E_1,p=\pm 1|z^\pm_\pm>=\left(z^+_\pm z^-_\pm\right)\,
e^{-\frac{1}{2}\left[|z^+_+|^2+|z^+_-|^2+|z^-_+|^2+|z^-_-|^2\right]}\ \ \ ,\\ \\
& &\ \ \ 
<E_1,p=0,\sigma=\pm 1|z^\pm_\pm>=
\frac{1}{\sqrt{2}}\left(z^+_+z^-_- \pm z^+_-z^-_+\right)\,
e^{-\frac{1}{2}\left[|z^+_+|^2+|z^+_-|^2+|z^-_+|^2+|z^-_-|^2\right]}\ \ \ ,
\end{array}
\end{equation}
corresponding to the Fock states
\begin{equation}
\begin{array}{r c l}
E_0=2\hbar g\omega\ &:&\ \ \ 
|E_0,p=0>=|0>\ \ \ ,\\ \\
E_1=4\hbar g\omega\ &:&\ \ \ 
|E_1,p=\pm 1>={\alpha^+_\pm}^\dagger\,{\alpha^-_\pm}^\dagger\,|0>\ \ \ ,\\ \\
& &\ \ \ 
|E_1,p=0,\sigma=\pm 1>=\frac{1}{\sqrt{2}}\,
\left({\alpha^+_+}^\dagger\,{\alpha^-_-}^\dagger\,\pm\,
{\alpha^+_-}^\dagger\,{\alpha^-_+}^\dagger\right)\,|0>\ \ \ ,
\end{array}
\end{equation}
where $p=0,\pm 1$ is the SO($d=2$) helicity label appearing in
(\ref{eq:partitiond2}), while the additional index $\sigma=\pm 1$
distinguishes the permutation properties under the exchange in the
indices $i=1$ and $i=2$ of the otherwise two degenerate states with $p=0$. 
Clearly, similar considerations apply to all levels of the physical energy 
spectrum, thus including representations of the permutation symmetry group
in the index $i=1,2$ to distinguish otherwise degenerate SO($d=2$)=U(1) 
representations.

Given the helicity coherent wave functions of all physical
states, it is possible to also determine their position wave functions
for the original degrees of freedom $q^a_i$, which will involve only
gauge invariant combinations of these degrees of freedom. 
In the $d=2$ case, these wave functions could be compared to those 
determined in Ref.\cite{Pause} over modular space, which is parametrised 
in terms of specific gauge invariant
combinations of the original degrees of freedom $q^a_i$. 
Even though this worthwhile exercise is not attempted here, there is
no doubt that identical results will be derived, involving the
appropriate special functions\cite{Pause}. Nevertheless, note how,
through the physical projector $\proj\,$, the coherent state basis 
allows for simple polynomial wave functions for gauge invariant states
(up to the characteristic gaussian factor $e^{-|z|^2/2}$), 
in direct correspondence with the Fock states spanning the physical spectrum.

\section{The SO(3) Gauge Invariant Model}
\label{Sect8}

The next simplest generalisation of the SO(2) model considered
so far, involves dynamics invariant under the non abelian gauge 
symmetry group SO(3), determined from the Lagrange function 
of real degrees of freedom $q^a_i$ and $\lambda^a$ $(a=1,2,3;i=1,2,\cdots,d)$,
\begin{equation}
L=\frac{1}{2g^2}\left[\dot{q}^a_i+\epsilon^{abc}\lambda^bq^c_i\right]^2
-V(q^a_i)\ \ \ ,
\label{eq:LSO(3)}
\end{equation}
with
\begin{equation}
V(q^a_i)=\frac{1}{2}\omega^2\,q^a_iq^a_i\ \ \ .
\label{eq:V3}
\end{equation}
Here again, $g$ and $\omega$ are arbitrary parameters with appropriate 
physical dimensions, and $\epsilon^{abc}$ is the three dimensional 
antisymmetric tensor such that $\epsilon^{123}=+1$. As a matter of fact, 
$\lambda^a$ correspond to the single time component of the non abelian 
SO(3) gauge field transforming under the three dimensional adjoint 
representation of SO(3), as do the matter fields $q^a_i$,
while $\epsilon^{abc}$ are the structure coefficients
of the associated Lie algebra\footnote{With the generators in 
the adjoint representation given by $(T^a)^{bc}=-i\epsilon^{abc}$.}.

Consequently, given the choice of SO(3) invariant potential in (\ref{eq:V3}), 
the Lagrange function (\ref{eq:LSO(3)}) describes the motion of $d$
three dimensional spherically symmetric harmonic oscillators, whose
total angular momentum vector must identically vanish at all times.
In particular, SO(3) gauge transformations of the coordinates $q^a_i$
and of the gauge variables $\lambda^a$ are of the form
\begin{equation}
{q'}^a_i=U^{ab}q^b_i\ \ ,\ \
{\lambda'}^a=U^{ab}\lambda^b-\frac{1}{2}\epsilon^{abc}U^{bd}\dot{U}^{cd}\ \ ,
\label{eq:transfSO(3)}
\end{equation}
where $U^{ab}$ is the $3\times 3$ rotation matrix defined in 
(\ref{A2:SO(3)matrix}) of Appendix B, whose Euler angles $\psi$,
$\theta$ and $\phi$ are arbitrary time dependent functions.

Clearly, in this case as well, the gauge variables $\lambda^a$ may be gauged
away entirely, leading in the $\lambda^a=0$ gauge to equations of motion
which are those of $d$ three dimensional spherically symmetric
harmonic oscillators of identical angular frequencies, constrained 
to possess an identically vanishing total angular momentum vector.

Similarly, the Hamiltonian analysis of the classical system goes through
as in the SO(2) case. The first-class Hamiltonian $H_0$ and gauge 
constraints $\phi^a$ are given by
\begin{equation}
H_0=\frac{1}{2}g^2p^a_ip^a_i+\frac{1}{2}\omega^2q^a_iq^a_i\ \ \ ,\ \ \
\phi^a=\epsilon^{abc}q^b_ip^c_i\ \ \ ,
\end{equation}
where $p^a_i$ are the momenta conjugate to the coordinates $q^a_i$,
while the total fundamental Hamiltonian reads
\begin{equation}
H=H_0-\lambda^a\phi^a\ \ \ ,
\end{equation}
the gauge variables $\lambda^a$ being indeed (opposite to)
the Lagrange multipliers
for the first-class constraints $\phi^a$. The latter, corresponding to the
cartesian components of the total angular momentum vector, generate the SO(3)
Lie algebra
\begin{equation}
\{\phi^a,\phi^b\}=\epsilon^{abc}\phi^c\ \ \ ,
\end{equation}
leave the first-class Hamiltonian invariant, $\{H_0,\phi^a\}=0$,
and induce the following infinitesimal SO(3) gauge transformations
on phase space, including those of the Lagrange multipliers $\lambda^a$,
\begin{equation}
\delta_\eta q^a_i=\epsilon^{abc}\eta^bq^c_i\ \ ,\ \ 
\delta_\eta p^a_i=\epsilon^{abc}\eta^bp^c_i\ \ ,\ \ 
\delta_\eta\lambda^a=-\dot{\eta}^a+\epsilon^{abc}\eta^b\lambda^c\ \ \ ,
\end{equation}
given arbitrary time dependent (infinitesimal) functions $\eta^a$.
Again, these transformations exponentiate to finite gauge transformations
identical to those (\ref{eq:transfSO(3)}) of the Lagrangian formulation.

It should be clear how in the gauge $\lambda^a=0$, the Hamiltonian
equations of motion are simply,
\begin{equation}
\dot{q}^a_i=g^2p^a_i\ \ \ ,\ \ \ \dot{p}^a_i=-\omega^2q^a_i\ \ \ ,
\end{equation}{
whose solutions are subjected to the constraints,
\begin{equation}
\phi^a=\epsilon^{abc}q^b_ip^c_i=0\ \ \ .
\end{equation}

\section{The SO(3) Model: Gauge Invariant Partition Functions}
\label{Sect9}

The canonical quantisation of the SO(3) model is straightforward enough,
and details are not given since they are similar to those for the
SO(2) model. In particular, the excitation level operator
$\hat{N}={\alpha^a_i}^\dagger{\alpha^a_i}$ extends the range of the
index $a=1,2,3$ in terms of creation and annihilation operators
defined as previously, while the quantum first-class Hamiltonian 
$\hat{H}_0$ reads,
\begin{equation}
\hat{H}_0=\hbar g\omega\left[\hat{N}+\frac{3}{2}d\right]\ \ \ .
\end{equation}

As for the SO(2) model, the physical spectrum of gauge invariant
states, including their degeneracies, is obtained from the partition
function,
\begin{equation}
{\rm Tr}\,x^{\hat{N}}\proj\,\ \ \ ,
\end{equation}
where $x$ is a complex parameter, possibly constrained by
$|x|<1$, and $\proj\,$ is the physical projection operator. In the
present model, the latter operator is defined by
\begin{equation}
\proj\,=\int_{\rm SO(3)}d\mu(\theta^a)
e^{\frac{i}{\hbar}\theta^a\hat{\phi}^a}=
\frac{1}{8\pi^2}\int_0^{2\pi}d\psi\int_0^\pi d\theta\,\sin\theta
\int_0^{2\pi}d\phi\,e^{\frac{i}{\hbar}\psi\hat{\phi}^3}
e^{\frac{i}{\hbar}\theta\hat{\phi}^1}
e^{\frac{i}{\hbar}\phi\hat{\phi}^3}\ \ \ ,
\end{equation}
where $\hat{\phi}^a=\epsilon^{abc}\hat{q}^b_i\hat{p}^c_i$ 
are the SO(3) generators of the quantised system. Indeed, in terms of the
Euler angle parametrisation given in Appendix B, the integration measure
$d\mu(\theta^a)$ is the SO(3) invariant Haar measure, 
while the product of transformations 
$e^{\frac{i}{\hbar}\psi\hat{\phi}^3}
e^{\frac{i}{\hbar}\theta\hat{\phi}^1}
e^{\frac{i}{\hbar}\phi\hat{\phi}^3}$ gives the corresponding SO(3)
transformation, which could also be expressed as 
$e^{\frac{i}{\hbar}\theta^a\hat{\phi}^a}$ in terms of specific
angles $\theta^a$ functions of the Euler angles $\psi$, $\theta$
and $\phi$.

Based on the arguments developed for the SO(2) model using coherent
states, it should be clear that the above partition function for the SO(3)
model is simply given by\footnote{The expression of the relevant determinant 
is given in Appendix B.}
\begin{equation}
{\rm Tr}\,x^{\hat{N}}\proj\,=\int_{\rm SO(3)}d\mu(\theta^a)\ \frac{1}
{\Big[{\rm det}\left(\delta^{ab}-xU^{ab}(\theta^a)\right)\Big]^d}\ \ \ ,
\label{eq:generalpartf}
\end{equation}
whose explicit evaluation leads to the following results\footnote{The
case $d=1$ needs to be considered separately from $d\ge 2$. The integration
over one of the two angles $\psi$ or $\phi$ is trivial since they
are summed, $\psi+\phi$, while the integration over the remaining one,
say $\phi$, is best done through the change of variable
$u={\rm tan}(\phi/2)$.}.

When $d=1$, one finds again
\begin{equation}
{\rm Tr}\,x^{\hat{N}}\proj\,=\sum_{n=0}^\infty x^{2n}=\frac{1}{1-x^2}\ \ \ .
\end{equation}
Thus in this case, the physical energy spectrum is given by
\begin{equation}
E_n=\hbar g\omega\left(2n+\frac{3}{2}\right)\ \ \ ,\ \ \ n=0,1,2,\dots\ \ \ ,
\end{equation}
and is free of any degeneracy at any level $n$,
\begin{equation}
d_n=1\ \ \ ,\ \ \ n=0,1,2,\dots\ \ \ .
\end{equation}

When $d\ge 2$, one has
\begin{equation}
{\rm Tr}\,x^{\hat{N}}\proj\,=\frac{(1+x)^{d-2}}{(1-x^2)^{3(d-1)}}\,
\sum_{n=0}^{d-2}\,\frac{(d-2)!}{n!\ (d-2-n)!}\,
\frac{(2n)!}{n!\ (n+1)!}\,x^n(1-x)^{2(d-2-n)}\ \ \ ,
\end{equation}
thus clearly displaying the existence of degeneracies at all
non trivial physical energy levels, starting with the first gauge invariant
excitation level $\hat{N}=2$ or $n=1$ for which $d_{n=1}=\frac{1}{2}d(d+1)$. 
For example for $d=2$, one finds results which of course agree with those
of Ref.\cite{Pause},
\begin{equation}
{\rm Tr}\,x^{\hat{N}}\proj\,=\frac{1}{(1-x^2)^3}=
\sum_{n=0}^\infty\,\frac{1}{2}(n+1)(n+2)\,x^{2n}\ \ \ ,
\end{equation}
showing that the physical energy spectrum is then given by,
\begin{equation}
E_n=\hbar g\omega\left(2n+3\right)\ \ \ ,\ \ \ n=0,1,2,\dots\ \ \ ,
\end{equation}
with the following degeneracies at each level $n$,
\begin{equation}
d_n=\frac{1}{2}(n+1)(n+2)\ \ \ ,\ \ \ n=0,1,2,\dots\ \ \ .
\end{equation}

In the case $d=3$, one finds
\begin{equation}
{\rm Tr}\,x^{\hat{N}}\proj\,=\frac{(1+x^3)}{(1-x^2)^6}=
\sum_{n=0}^\infty\,\frac{(n+5)!}{5!\ n!}\,x^{2n}(1+x^3)\ \ \ ,
\end{equation}
establishing that the physical energy spectrum is separated into
two classes with iden\-ti\-cal de\-ge\-ne\-ra\-cies,
for all values $n=0,1,2,\dots$,
\begin{equation}
\begin{array}{r c l}
E^{(1)}_n=\hbar g\omega\left(2n+\frac{9}{2}\right)\ \ \ &,&\ \ \ 
d^{(1)}_n=\frac{1}{120}(n+1)(n+2)(n+3)(n+4)(n+5)\ \ \ ,\\ \\
E^{(2)}_n=\hbar g\omega\left(2n+\frac{15}{2}\right)\ \ \ &,&\ \ \ 
d^{(2)}_n=\frac{1}{120}(n+1)(n+2)(n+3)(n+4)(n+5)\ \ \ .
\end{array}
\end{equation}
The appearance of this new feature is related to the fact that
starting with $d=3$, new SO(3) gauge invariant combinations of the 
creation operators ${\alpha^a_i}^\dagger$ become possible using 
the SO(3) invariant structure coefficients $\epsilon^{abc}$, 
in addition to the only other SO(3) invariant tensor 
$\delta^{ab}$ which may be used whatever the value of $d$.

Similarly, for still larger values of $d$, the structure of the physical
energy spectrum and its degeneracies becomes even richer and still more
involved, in a manner which deserves to be fully understood. In particular,
the SO($d$) global symmetry of the model goes some way in accounting
for the observed degeneracies, but here again, some larger symmetry
still to be identified must be at work. Considering nevertheless
the SO($d$) Cartan subalgebra valued partition function, it should be
clear that this quantity is given by the following integral over the
SO(3) manifold,
\begin{equation}
{\rm Tr}\,x^{\hat{N}}e^{-\frac{1}{2}\omega_{ij}\hat{L}_{ij}}\proj\,=
\int_{\rm SO(3)}d\mu(\theta^a)\,\frac{1}
{{\rm det}\left[\delta^{ab}\delta_{ij}-xU^{ab}(\theta^a)
R_{ij}(\omega_{ij})\right]}
\ \ \ ,
\end{equation}
with the same notations and operators as those introduced already for
the SO(2) model. Refraining from establishing here general results for
arbitrary values of $d$, let us concentrate on the $d=2$ case, which
carries with it the basic structure for a general solution as was
already observed for the SO(2) model. Working in the helicity basis
for the index $i=1,2$, an explicit calculation then finds
\begin{equation}
{\rm Tr}\,x^{\hat{N}}e^{-\frac{i}{\hbar}\omega_{ij}\hat{L}_{ij}}\proj\,=
\frac{1}{(1-x^2)(1-x^2e^{2i\omega_{12}})(1-x^2e^{-2i\omega_{12}})}=
\sum_{n=0}^\infty x^{2n}\sum_{p=-n}^{n}d(n,p)e^{2ip\omega_{12}}\ \ \ ,
\end{equation}
with the SO($d=2$) helicity degeneracies
\begin{equation}
d(n,p)=E\left[\frac{n+|p|}{2}\right]-|p|+1\ \ \ ,\ \ \ 
-n\le p\le n\ \ ,\ \ n=0,1,2,\dots\ \ \ ,
\end{equation}
where $E[\rho]$ is the integer part of a real variable $\rho$, 
such that indeed,
\begin{equation}
\sum_{p=-n}^{n}d(n,p)=\frac{1}{2}(n+1)(n+2)\ \ \ ,\ \ \ n=0,1,2,\dots\ \ \ .
\end{equation}

Even though possible at this stage, the extension of these results to 
larger values of $d$ is to be addressed in later work, which should also 
include the construction of the generating function for coherent state wave 
functions of all gauge invariant physical states. 
Indeed, once identified, and with the help of the appropriate
group valued partition and generating functions for physical states,
the complete symmetry accounting for all
degeneracies of the physical spectrum, including those of the SO($d$) 
representations themselves, should in principle render the analysis of
all these issues more transparent. This specific aspect of this class
of models is deferred to later work. Nevertheless, the analysis of the
present paper has clearly demonstrated the advantages in using the physical
projection operator in conjunction with techniques of coherent states, 
to disentangle the physical spectrum of gauge invariant systems.

\section{Conclusions}
\label{Sect10}

In this paper, the physical projector operator constructed in
Ref.\cite{Klauder1} for general constrained systems, is applied
to a particular class of gauge invariant quantum mechanical systems.
In contradistinction to more standard approaches which
require gauge fixing and generally also additional degrees of freedom,
by working only with the initial variables of a system,
the physical projector approach to gauge invariant systems avoids
these requirements altogether, and thereby also the ensuing
delicate issues of possible Gribov problems and the characterization of
modular space. Nevertheless, it provides
both a manifestly covariant as well as a consistent unitary description 
of the dynamics of gauge invariant physical states. As the present analysis 
has demonstrated even within the context of relatively simple models, 
when conjugated with techniques of coherent states, 
the physical projector approach offers many advantages over
more traditional ones. For instance, at least for the models considered, it
reduces the determination of their physical spectrum to an exercise 
in the combinatorics of compact Lie group representation theory
and in the integration over group manifolds of characters of these 
representations\footnote{Indeed, in the general case, the evaluation of 
a quantity such as (\ref{eq:generalpartf}), and more specifically of 
the associated generating function of physical state wave functions, 
involves precisely the group integration of characters of representations, 
when expanding these expressions in terms of traces of powers of 
the matrices involved.}.

Even though only two types of models were considered
explicitly in the paper, namely one with the abelian gauge group SO(2) 
and the other with the non abelian gauge group SO(3), these examples 
belong to quite general classes of systems whose dynamics is described 
by Lagrange functions of the form\footnote{Such Lagrange function may 
be extended to arbitrary complex representations of the gauge group as well,
in an obvious way.},
\begin{equation}
L=\frac{1}{2g^2}\left[\dot{q}^\alpha_i+i\lambda^a
\left(T^a\right)^{\alpha\beta}q^\beta_i\right]^2-V(q^\alpha_i)\ \ \ .
\label{eq:generalL}
\end{equation}
Here, $q^\alpha_i(t)$ $(\alpha=1,2,\cdots,r;\ i=1,2,\cdots,d)$
are a collection of $d$ scalar matter fields transforming in some
$r$-dimensional real representation of a gauge group $G$, whose generators 
are (dim\ $G$) hermitian $r\times r$ matrices $T^a$ 
$(a=1,2,\cdots,{\rm dim}\ G)$.
The gauge field has as single time component the real quantities
$\lambda^a(t)$ $(a=1,2,\cdots,{\rm dim}\ G)$. Finally, $V(q^\alpha_i)$ is
some specific $G$-invariant potential for the matter fields $q^\alpha_i$.

The models considered in the present paper correspond to the gauge groups
SO(2) and SO(3), with the $d$ matter fields in the defining representation
of these groups, and a harmonic potential of identical angular frequency
for all $d$ matter fields. More generally, the gauge group SO($N$) with
$d$ matter fields in the $N$-dimensional representation could be 
considered\footnote{The case $d=1$ having already been discussed in 
Refs.\cite{Shabanov1,Shabanov2}, whose results agree of course with those
of the present analysis for SO($N=2$) and SO($N=3$) when $d=1$.}. 
Another possibility is to choose the $d$ matter fields in the 
(dim\ $G$)-dimensional adjoint representation of the gauge group, 
in which case the matrices $T^a$ are given by the structure coefficients 
$f^{abc}$ of the associated Lie algebra, $(T^a)^{bc}=-if^{abc}$. 
Note that the $G=\,$SO(3) model of the paper belongs to this class, 
while for an arbitrary gauge group $G$, the case with $d=1$ has already 
been considered in Refs.\cite{Shabanov2,Gov2}.

Clearly, the approach of the present paper may be applied to many classes 
of such models, including those based on the exceptional
Lie algebras (starting for example with $G_2$ whose representations are 
all real). Beyond those cases, the analysis of 1+1 dimensional
non abelian pure Yang-Mills theories (possibly also coupled to bosonic 
or fermionic matter fields) may also be addressed along similar lines, by
compactifying, for example, the space dimension onto a circle (this type
of model is also considered in Ref.\cite{Pause} for SO(3)). Indeed, 
the dimensional reduction of such models to 0+1 dimensions belongs precisely 
to the general class of systems defined by (\ref{eq:generalL}). Likewise, 
the compactifications to 1+1 and 0+1 dimensions of models motivated by 
the dualities of $M$-theory\cite{Mtheory} could provide another fertile 
field for the physical projector approach. Finally, the potentialities 
offered by this approach are there to be explored beyond 1+1 dimensional 
Yang-Mills systems, beginning with the fascinating case of 2+1 dimensional 
Chern-Simons theories\cite{Witten1}, both for compact and non-compact
Lie groups, the latter case being related to theories of gravity\cite{Witten2}.

The general programme just outlined should also help unravel, and
benefit from, those issues left open in the present work, related to 
the genuine reasons for the observed degeneracies in the physical energy 
spectrum of the SO(2) and SO(3) models. As explained in Appendix A
in the case of the $d$-dimensional spherical harmonic oscillator, the 
classical symmetries of a system may be enhanced at the quantum level into
dynamical symmetries. Indeed, for the general class of models defined
by (\ref{eq:generalL}), when the potential $V(q^\alpha_i)$
is chosen to be that of $r\times d$
harmonic oscillators of identical angular frequency $\omega$,
the quantised system possesses a {\sl global\/} dynamical symmetry based 
on the unitary group U($r d$)=U(1)$_N\times$SU($r d)$ acting on the
matter fields $q^\alpha_i$, independently of the existence of the 
local gauge symmetries based on the group $G$. The U(1)$_N$ symmetry is 
generated by the total excitation level operator $\hat{N}$, and thus acts 
in a trivial manner on the space of states. On the other hand, the SU($r d$)
symmetry possesses the subgroup SU($r$)$\times$SU($d$), where the first
factor acts on the $\alpha$ indices of the coordinates $q^\alpha_i$, while
the second factor acts on the $i$ indices. The dynamical symmetry SU($d$) 
thus enhances at the quantum level the global SO($d$) symmetry of the 
classical system, the latter being embedded into the former. Similarly,
the {\sl local\/} gauge symmetry group $G$ is a subgroup of the
{\sl global\/} dynamical symmetry SU($r$), and it is by understanding
how $G$ is embedded into SU($r$) that the global SU($d$) symmetry 
properties of gauge invariant physical states can be determined.

Indeed, since the creation operators ${\alpha^\alpha_i}^\dagger$ all
commute with one another and transform in the fundamental $rd$-dimensional
representation of SU($rd$), at the excitation level $\hat{N}=n$, all 
states of the quantised system fall into the $n$ times totally symmetric
representation of SU($rd$). Reducing that representation to the
subgroups SU($r$)$\times$SU($d$) and then $G\times$SU($d$), and
retaining only those representations which are singlets under the 
gauge symmetry group $G$, leads to the identification of all physical 
states at excitation level $\hat{N}=n$ including their SU($d$) symmetry 
properties. Consequently, the symmetry accounting
for all degeneracies in the physical energy spectrum is the dynamical 
group\footnote{When the variables $q^\alpha_i$ transform under a 
complex representation of the gauge group $G$, this conclusion must 
be adapted appropriatedly.} SU($d$). Hence, rather than the classical 
SO($d$) symmetry, it is the quantum dynamical SU($d$) symmetry, through 
its Cartan sublagebra, which must be used in the partition and wave 
function generating functions to ``tag" physical states and their 
global symmetry properties, possibly also including a ``tagging" for 
the SU($d$) Casimir operators in order to easily disentangle the 
SU($d$) representation content of the physical 
spectrum, and in particular the wave functions of the highest weight states 
and their descendants in each of the SU($d$) representations. By combining 
the insight provided by this dynamical symmetry with the advantages of the 
physical projector and coherent state techniques, it thus appears possible 
to give a completely structured characterization of the physical spectrum of
the class of models defined by (\ref{eq:generalL}).

In the case of the SO(2) and SO(3) models of the present paper,
all physical states thus fall into representations of SU($d$).
For example, at the first non trivial physical excitation level
$\hat{N}=2$, these representations are as follows. When $G=\,$SO(2),
one finds the two index symmetric and antisymmetric SU($d$) representations,
of dimensions $d(d+1)/2$ and $d(d-1)/2$ respectively, hence indeed a total
of $d^2$ states. When $G=\,$SO(3), one finds only the two index symmetric
SU($d$) representation, indeed a total of $d(d+1)/2$ states.

The dynamical symmetry also helps disentangle the wave functions from
the generating function. To illustrate this point, consider simply
the SO(2) model with $d=2$, for which the generating function
is given in (\ref{eq:wavefSO(2)}). How is one to associate
the $(n+1)(n+2)(n+3)/6$ terms stemming from the $n$-th power of the
sum of four terms to the $(n+1)^2$ physical states at the excitation 
level $\hat{N}=2n$? The answer to this question is not immediate
from the helicity quantum number $-n\le p\le n$ associated to the
global SO($d=2$) symmetry, but it is from the 
SU($d=2$) dy\-na\-mi\-cal symmetry\footnote{Note that the dynamical SU($d$)
symmetry also includes the permutation symmetry group $S_d$ acting on the
$d$ indices $i$, a discrete symmetry which was used to lift 
degeneracies in the SO(2) case with $d=2$ (see Sect.\ref{Sect7}).}. 
The global SU(4) dynamical symmetry in this case
being of rank 3, its Cartan sublagebra is spanned by that of the subgroup
SU($r=2$)$\times$SU($d=2$), which is of rank 2, and an intertwining
operator $\hat{S}$ coupling these two SU(2) factors. Moreover, the
embedding of the gauge group SO(2) and of the global symmetry group
SO($d=2$) into each of these SU(2) factors may be chosen such that
their generators $\hat{\phi}$ and $\hat{L}_{12}$ coincide with
the respective Cartan subalgebra $T_3$ generators both in SU($r=2$) 
and in SU($d=2$).  Consequently, physical states are such that their $T_3$ 
eigenvalue for SU($r=2$) vanishes while their $T_3$ eigenvalue for SU($d=2$)
corresponds to the SO($d=2$) helicity quantum number $p$.
Thus in addition to their energy (or excitation level) eigenvalue,
physical states are labelled by SU($d=2$) quantum numbers, namely
a ``spin" value $T$ as well as the $T_3=p$ projection in SU($d=2$),
{\sl i.e.} $|E_n,T,T_3>$. At physical energy level $E_n$, the ``spin"
content of states is $T=0,1,2,\cdots,n$, with each of these representations
appearing only once, establishing that the dynamical SU($d=2$) symmetry
does indeed account completely for all degeneracies in the physical
spectrum of the model. And the same conclusion also enables the correct
identification of the wave functions for all states from the generating 
function (\ref{eq:wavefSO(2)}), including their normalisation.
At energy level $E_n$, one way is to start from the highest spin state
with $T=n$ and $T_3=n$, and using the SU($d=2$) lowering operator,
identify the wave functions of all other states at that energy level
having the same ``spin" value $T$ but different values of $T_3=p$. 
Repeating the same procedure over and over again, starting each time
from the remaining highest spin state, the whole
content of states at energy level $E_n$ can be exhausted that way.
However, a more efficient approach would be to introduce into the partition
and generating functions a ``tagging" in terms of the SU($d=2$)
Casimir operator, thereby immediately associating their $T$ and $T_3$
values to each of the physical states and their wave functions.

Clearly, the same considerations using dynamical symmetries in
conjunction with the physical projector operator and coherent states, 
apply to the general classes of gauge invariant quantum mechanical
models defined by (\ref{eq:generalL}), and beyond those, 
to the genuine gauge invariant field theories mentioned previously.
Progress along these lines is left for future work.

\section*{Acknowledgements}

J. G. gratefully acknowledges the warm and generous hospitality of
the members and of the Department of Physics at the University of Florida, 
were most of this work was completed.

\clearpage

\section*{Appendix A}

Consider the ordinary one dimensional harmonic oscillator,
of mass $m$ and angular frequency $\omega$. Its quantum Hamiltonian
$\hat{H}$,
\begin{equation}
\hat{H}=\frac{\hat{p}^2}{2m}+\frac{1}{2}m\omega^2\hat{q}^2=
\hbar\omega(a^\dagger a+\frac{1}{2})\ \ \ ,
\end{equation}
is given in terms of the pairs of conjugate degrees of freedom,
either the (position, momentum) operators $\hat{q}$ and $\hat{p}$,
or the (annihilation, creation) operators $a$ and $a^\dagger$, obeying
the usual algebraic relations
\begin{equation}
\hat{q}^\dagger=\hat{q}\ \ ,\ \ \hat{p}^\dagger=\hat{p}\ \ ,\ \ 
[\hat{q},\hat{p}]=i\hbar\ \ \ ,\ \ \ [a,a^\dagger]=1\ \ .
\end{equation}
The relations between these different operators are
\begin{equation}
a=\sqrt{\frac{m\omega}{2\hbar}}\,\left[\hat{q}+
i\frac{\hat{p}}{m\omega}\right]\ \ \ ,\ \ \ 
a^\dagger=\sqrt{\frac{m\omega}{2\hbar}}\,\left[\hat{q}-
i\frac{\hat{p}}{m\omega}\right]\ \ \ ,\ \ \ 
\label{eq:aq}
\end{equation}
\begin{equation}
\hat{q}=\sqrt{\frac{\hbar}{2m\omega}}\,\left[a+a^\dagger\right]\ \ \ ,\ \ \ 
\hat{p}=-i\sqrt{\frac{m\hbar\omega}{2}}\,\left[a-a^\dagger\right]\ \ \ .
\label{eq:qa}
\end{equation}

As is well known, the Fock space representation of the $(a,a^\dagger)$ algebra
is spanned by a basis of orthonormalised states $|n>$ ($n=0,1,\dots$) 
which also diagonalise the quantum Hamiltonian $\hat{H}$,
\begin{equation}
|n>=\frac{1}{\sqrt{n!}}\,(a^\dagger)^n\,|0>\ \ \ ,\ \ \ 
\hat{H}|n>=\hbar\omega(n+\frac{1}{2})|n>\ \ \ ,\ \ \ n=0,1,2,\dots\ \ \ ,
\end{equation}
$|0>$ being the normalised vacuum state annihilated by the operator $a$,
while the states $|n>$ are thus such that $<n|m>=\delta_{nm}$.

By analogy with (\ref{eq:aq}) and (\ref{eq:qa}), and given real values
$q$ and $p$ related to the phase space degrees of freedom of the system,
let us introduce the complex quantities
\begin{equation}
z=\sqrt{\frac{m\omega}{2\hbar}}\,\left[q+i\frac{p}{m\omega}\right]\ \ \ ,\ \ \ 
\bar{z}=\sqrt{\frac{m\omega}{2\hbar}}\,\left[q-i\frac{p}{m\omega}\right]\ \ \ ,
\label{eq:zq}
\end{equation}
so that,
\begin{equation}
q=\sqrt{\frac{\hbar}{2m\omega}}\,\left[z+\bar{z}\right]\ \ \ ,\ \ \ 
p=-i\sqrt{\frac{m\hbar\omega}{2}}\,\left[z-\bar{z}\right]\ \ \ .
\label{eq:qz}
\end{equation}
Associated to these quantities, one introduces the holomorphic
and phase space coherent states, defined, respectively, by
\begin{equation}
|z>=e^{-\frac{1}{2}|z|^2}\,e^{za^\dagger}|0>\ \ \ ,\ \ \
|p,q>=e^{-\frac{i}{\hbar}q\hat{p}}\,e^{\frac{i}{\hbar}p\hat{q}}\,|0>\ \ \ .
\end{equation}
Given the identity,
\begin{equation}
za^\dagger-\bar{z}a=-\frac{i}{\hbar}q\hat{p}+\frac{i}{\hbar}p\hat{q}\ \ \ ,
\end{equation}
these two sets of coherent states are essentially equivalent up to
a phase factor, as follows,
\begin{equation}
|z>=e^{\frac{i}{2\hbar}qp}\,|p,q>\ \ \ ,
\end{equation}
provided of course the variables $z$, $q$ and $p$ are related as in
(\ref{eq:zq}) and (\ref{eq:qz}).

The interest of these coherent states lies with the fact that they
also generate the whole representation space of the quantum oscillator.
Indeed, the identity operator possesses the following resolutions,
\begin{equation}
\one=\sum_{n=0}^\infty\,|n><n|=\int\frac{dzd\bar{z}}{\pi}\,|z><z|=
\int_{(\infty)}\frac{dqdp}{2\pi\hbar}\,|p,q><p,q|\ \ \ .
\end{equation}
Consequently, even though coherent states provide an overcomplete
basis of states, they allow for more straightforward calculations.
Indeed, quantum wave functions often reduce to simple polynomials
in the variables $z$ or $\bar{z}$, rather than special functions
for the ordinary position or momentum wave function representations.
Thus, the basic matrix elements of direct use are simply
\begin{equation}
a^n|z>=z^n|z>\ \ \ ,\ \ \ 
<n|z>=\frac{1}{\sqrt{n!}}\,z^n\,e^{-\frac{1}{2}|z|^2}\ \ \ ,\ \ \ 
n=0,1,2,\dots\ \ \ ,
\label{eq:nz}
\end{equation}
while one also has
\begin{equation}
<z_1|z_2>=e^{-\frac{1}{2}|z_1|^2-\frac{1}{2}|z_2|^2+\bar{z_1}z_2}\ \ \ ,\ \ \
<z|z>=1\ \ \ ,
\end{equation}
or equivalently
\begin{equation}
<p_2,q_2|p_1,q_1>=e^{-\frac{m\omega}{4\hbar}
\left[(q_2-q_1)^2+\left(\frac{p_2-p_1}{m\omega}\right)^2
-\frac{2i}{m\omega}(p_2+p_1)(q_2-q_1)\right]}\ \ \ ,\ \ \ 
<p,q|p,q>=1\ \ \ .
\end{equation}
These expressions should be contrasted, for example, to the position
wave functions of the energy eigenstates $|n>$,
\begin{equation}
<q|n>=\left(\frac{m\omega}{\pi\hbar}\right)^{\frac{1}{4}}\,
\frac{1}{\sqrt{2^n n!}}\,e^{-\frac{m\omega}{2\hbar}q^2}\,
H_n\left(q\sqrt{\frac{m\omega}{\hbar}}\right)\ \ \ ,\ \ \ n=0,1,2,\dots\ \ \ ,
\end{equation}
$H_n(x)$ being the usual Hermite polynomials, and $|q>$ the orthonormalised 
basis of position eigenstates such that $<q|q'>=\delta(q-q')$.

Clearly, given the different results above, it is possible to construct
the position wave function representation $<q|\psi>$ of any state $|\psi>$
from its coherent state wave function $<z|\psi>$ which typically, up to
the gaussian factor $e^{-\frac{1}{2}|z|^2}$, is simply some polynomial 
in $\bar{z}$ (see (\ref{eq:nz}) for an example).

Consider now two such harmonic oscillators, labelled by the indices
$i=1,2$, with identical mass $m$ and angular frequency $\omega$, or
in other words, a two dimensional harmonic oscillator with circular
symmetry. All the above discussion goes through in that case in an obvious
way, since the corresponding representation space is simply the tensor
product, over the index values $i=1$ and $i=2$, of the previous one.
Nevertheless, in order to make explicit the global SO(2)=U(1) circular
symmetry of the system, it proves efficient to work, rather than with
the cartesian basis $(q_1,q_2)$, with a helicity-like basis of coordinates
$q_{\pm}=(q_1\mp iq_2)/\sqrt{2}$. Indeed, SO(2) rotations of the cartesian
coordinates by an angle $\theta$ then correspond to
U(1) phase transformations $e^{\pm i\theta}$ of the helicity coordinates.

More specifically, rather than working with the usual annihilation
and creation operator algebra $(a_i,a^\dagger_i)$ ($i=1,2$), let us introduce
the quantities
\begin{equation}
a_\pm=\frac{1}{\sqrt{2}}\left[a_1\mp ia_2\right]\ \ \ ,\ \ \ 
a_\pm^\dagger=\frac{1}{\sqrt{2}}\left[a_1^\dagger\pm ia_2^\dagger\right]\ \ \ ,
\end{equation}
and likewise, given complex variables $z_i$ $(i=1,2)$ related to
the cartesian phase space coordinates $(q_i,p_i)$ $(i=1,2)$ in the same
manner as above, consider the combinations
\begin{equation}
z_\pm=\frac{1}{\sqrt{2}}\left[z_1\mp iz_2\right]\ \ \ ,\ \ \ 
\bar{z}_\pm=\frac{1}{\sqrt{2}}\left[\bar{z}_1\pm i\bar{z}_2\right]\ \ \ .
\end{equation}
Quite obviously, the ensuing helicity basis quantum algebra is still
given by two sets of commuting creation and annihilation operators,
\begin{equation}
[a_+,a_-^\dagger]=0=[a_-,a_+^\dagger]\ \ \ ,\ \ \
[a_+,a_+^\dagger]=1=[a_-,a_-^\dagger]\ \ \ ,
\end{equation}
while the following relations also hold
\begin{equation}
z_1a_1^\dagger+z_2a_2^\dagger=z_+a_+^\dagger+z_-a_-^\dagger\ \ \ ,\ \ \ 
|z_1|^2+|z_2|^2=|z_+|^2+|z_-|^2\ \ \ .
\end{equation}

The quantum Hamiltonian then reads,
\begin{equation}
\hat{H}=\hbar\omega\left[a_1^\dagger a_1+a_2^\dagger a_2+1\right]=
\hbar\omega\left[a_+^\dagger a_++a_-^\dagger a_-+1\right]\ \ \ ,
\label{eq:H2}
\end{equation}
as well as the generator of global SO(2)=U(1) symmetry transformations,
\begin{equation}
Q=-i\left[a_1^\dagger a_2-a_2^\dagger a_1\right]=
a_+^\dagger a_+-a_-^\dagger a_-\ \ \ ,
\end{equation}
such that,
\begin{equation}
[Q,a_\pm]=\mp a_\pm\ \ \ ,\ \ \ 
[Q,a_\pm^\dagger ]=\pm a_\pm^\dagger \ \ \ .
\end{equation}
Correspondingly, the Hamiltonian is diagonalised by the helicity Fock
space basis,
\begin{equation}
|n_+,n_->=\frac{1}{\sqrt{n_+!\ n_-!}}\left(a_+^\dagger\right)^{n_+}
\left(a_-^\dagger\right)^{n_-}|0>\ \ ,\ \ n_+,n_-=0,1,2,\dots\ \ ,
\label{eq:basis2}
\end{equation}
\begin{equation}
\hat{H}|n_+,n_->=E(n_+,n_-)|n_+,n_->\ \ ,\ \ 
E(n_+,n_-)=\hbar\omega(n_++n_-+1)|n_+,n_->\ \ ,
\label{eq:H2energy}
\end{equation}
while the helicity (holomorphic) coherent states are defined by,
\begin{equation}
|z_+,z_->=e^{-\frac{1}{2}\left[|z_+|^2+|z_-|^2\right]}\,
e^{z_+a_+^\dagger}\,e^{z_-a_-^\dagger}\,|0>\ \ \ .
\end{equation}
In terms of these generating vectors, the resolution of the identity
is expressed as,
\begin{equation}
\one=\sum_{n_+,n_-=0}^{\infty}|n_+,n_-><n_+,n_-|=
\int\prod_{\pm}\frac{dz_\pm d\bar{z}_\pm}{\pi}\,|z_+,z_-><z_+,z_-|\ \ \ ,
\label{eq:unity}
\end{equation}
with the basic matrix elements for the change of basis given by,
\begin{equation}
<n_+,n_-|z_+,z_->=\frac{1}{\sqrt{n_+!n_-!}}\,z_+^{n_+}\,z_-^{n_-}\,
e^{-\frac{1}{2}\left[|z_+|^2+|z_-|^2\right]}\ \ \ .
\label{eq:helicityFockbasis}
\end{equation}

It also proves interesting to consider the action of the SO(2)=U(1) rotation
generator $Q$ on the coherent states. A straightforward analysis
using (\ref{eq:unity}) and (\ref{eq:helicityFockbasis}) finds,
\begin{equation}
e^{i\theta Q}|z_+,z_->=|e^{i\theta}z_+,e^{-i\theta}z_->\ \ \ ,
\end{equation}
thus confirming the above claim as to the property of the helicity
basis under SO(2)=U(1) transformations. In terms of the cartesian
variables $z_i$ $(i=1,2)$, the phase transformations 
$z'_\pm=e^{\pm i\theta}z_\pm$
correspond to the SO(2) rotation,
\begin{equation}
\left(\begin{array}{c}
	z'_1\\
	z'_2
      \end{array}\right)=
\left(\begin{array}{c c}
	\cos\theta & \sin\theta\\
	-\sin\theta & \cos\theta
      \end{array}\right)\
\left(\begin{array}{c}
	z_1\\
	z_2
      \end{array}\right)\ \ \ ,
\end{equation}
or equivalently in matrix notation, to $z'_i=U_{ij}(\theta)z_j$ $(i=1,2)$
with the rotation matrix,
\begin{equation}
\left[U_{ij}(\theta)\right]=
\left(\begin{array}{c c}
	\cos\theta & \sin\theta \\
	-\sin\theta & \cos\theta
      \end{array}\right)\ \ \ .
\end{equation}
Consequently, the action of the rotation generator on the coherent
states $|z_i>$ in cartesian coordinates is given by,
\begin{equation}
e^{i\theta Q}|z_i>=|U_{ij}(\theta)z_j>\ \ \ .
\label{eq:rotation}
\end{equation}

As a matter of fact, the two dimensional system possesses a global
symmetry larger than the SO(2)=U(1) rotational one considered so far.
Indeed, even though the {\em classical\/} harmonic oscillator
is invariant under the global SO(2)=U(1) transformations constructed
above, {\em at the quantum level\/}, it does possess a {\em dynamical\/}
global unitary U(2)=SU(2)$\times$ U(1)$_N$ symmetry extending the initial SO(2) 
one. This U(2) symmetry is related to the possibility, {\em appearing
only at the quantum level\/}, to perform arbitrary U(2) rotations among 
the creation (or the annihilation) operators $a_i^\dagger$ 
(or $a_i$) ($i=1,2$), or $a_\pm^\dagger$ (or $a_\pm$), while the Hamiltonian 
$\hat{H}$ in (\ref{eq:H2})
is invariant under these unitary transformations. The generator associated
to the U(1)$_N$ factor of the U(2) symmetry is the excitation level
operator $\hat{N}$, given by
\begin{equation}
\hat{N}=a_1^\dagger a_1+a_2^\dagger a_2=a_+^\dagger a_++a_-^\dagger a_-\ \ \ .
\end{equation}
The expression of the remaining SU(2) generators in terms of the
creation and annihilation ope\-ra\-tors is defined up to an SU(2) 
homeomorphism,
namely up to an arbitrary unitary linear combination of the cartesian
or helicity coordinates. Necessarily, the generator $Q$ of the original
global SO(2)=U(1) symmetry is then also a certain specific linear
combination in the SU(2) algebra, thereby determining the SO(2) embedding
in SU(2).

The preferred choice of embedding is such that $Q$ coincides with 
the third generator $T^3$ of SU(2), and is best expressed in the helicity 
basis. Given the usual Pauli matrices $\tau^a$ ($a=1,2,3$),
\begin{equation}
\tau^1=
\left(\begin{array}{c c}
	0 & 1 \\
	1 & 0 
	\end{array}\right)\ \ ,\ \ 
\tau^2=
\left(\begin{array}{c c}
	0 & -i \\
	i & 0 
	\end{array}\right)\ \ ,\ \ 
\tau^3=
\left(\begin{array}{c c}
	1 & 0 \\
	0 & -1 
	\end{array}\right)\ \ ,
\end{equation}
a basis of SU(2) generators is specified by\footnote{A similar definition
is possible in terms of the cartesian creation and annihilation operators
$a_i^\dagger$ and $a_i$ $(i=1,2)$, in which case it is the generator $T^2$ 
which coincides with the SO(2)=U(1) charge $Q/2$.}
\begin{equation}
T^a=\sum_{\eta,\eta'=+,-}\,a^\dagger_{\eta'}
\left(\frac{\tau^a}{2}\right)_{\eta',\eta}\,a_\eta
=a^\dagger\cdot\frac{\tau^a}{2}\cdot a\ \ \ ,
\end{equation}
whose algebra is thus,
\begin{equation}
{T^a}^\dagger=T^a\ \ ,\ \ 
[T^a,T^b]=i\epsilon^{abc}T^c\ \ ,\ \ a,b,c=1,2,3\ \ \ .
\end{equation}
In particular, the raising and lowering operators $T^\pm=T^1\pm i T^2$
then reduce to
\begin{equation}
T^+=a_+^\dagger a_-\ \ \ ,\ \ \ T^-=a_-^\dagger a_+\ \ \ ,
\end{equation}
while $T^3$ then coincides indeed with the SO(2) rotation generator $Q/2$,
\begin{equation}
T^3=\frac{1}{2}\left[a_+^\dagger a_+-a_-^\dagger a_-\right]=\frac{1}{2}Q\ \ \ .
\end{equation}

The dynamical SU(2) symmetry in fact explains the degeneracies of the
energy spectrum $E(n_+,n_-)$ in (\ref{eq:H2energy}). For a fixed
value of $n=n_++n_-$, corresponding to states $|n_+,n_->$ sharing
a common energy value $\hbar\omega(n+1)$, all these states span
the $n$ times totally symmetric SU(2) representation of dimension
$(n+1)$, {\em i.e.\/} of ``spin" $n/2$, while they are distinguished by their
$Q$ or $2T^3$ eigenvalue $(n_+-n_-)$. Indeed, since the creation operators
$a_+^\dagger$ and $a_-^\dagger$ commute and define the fundamental SU(2) 
representation, the states (\ref{eq:basis2}) at level $\hat{N}=n=n_++n_-$ 
all fall into the representation of ``spin" $n/2$, and are related to one 
another by the action of the raising and lowering operators $T^+$ and $T^-$. 
Thus for example, the state $(a_-^\dagger)^n|0>/\sqrt{n!}$ defines the lowest
weight state of the representation of ``spin" $n/2$, of energy
$E=\hbar\omega(n+1)$ and of ``spin" projection $(-n/2)$, while all the 
other properly orthonormalised states of the same energy and of ``spin"
projection ranging from $(-(n-1)/2)$ up to $n/2$, are obtained by the repeated 
action---up to $n$ times---of the simple positive root $T^+$.

Such dynamical symmetries are characteristic of spherical harmonic oscillators
in any $d$ dimensional Euclidean space. Indeed, beyond the classical
SO($d$) symmetry acting on the cartesian coordinates parametrising such systems,
at the quantum level, U($d$)=SU($d$)$\times$U(1)$_N$ unitary transformations
among the cartesian creation (or annihilation) operators $a_i^\dagger$ 
($i=1,2,\cdots,d$) define a dynamical symmetry commuting with the quantum
Hamiltonian $\hat{H}$, which, up to a constant and a factor, coincides
with the excitation level operator $\hat{N}$ generating the U(1)$_N$
symmetry,
\begin{equation}
\hat{H}=\hbar\omega\left[\sum_{i=1}^d\,a_i^\dagger a_i\,+\,\frac{1}{2}d\right]=
\hbar\omega\left[\hat{N}+\frac{1}{2}d\right]\ \ \ .
\label{eq:Hd}
\end{equation}
Given the algebra
\begin{equation}
[a_i,a_j^\dagger]=\delta_{ij}\ \ \ ,\ \ \ i,j=1,2,\cdots,d\ \ \ ,
\end{equation}
holomorphic coherent states are defined in the same way as in the
one dimensional case, in terms of a collection of complex variables $z_i$
which may also be related to cartesian phase space coordinates
$q_i$ and $p_i$ ($i=1,2,\cdots,d$),
\begin{equation}
|z_i>=e^{-\frac{1}{2}|z_i|^2}e^{z_ia_i^\dagger}|0>\ \ \ .
\end{equation}

In order to define the action of the dynamical SU($d$) symmetry on these
coherent states, since the creation (or annihilation) operators
$a_i^\dagger$ transform in the fundamental SU($d$) representation of 
dimension $d$, let us choose a set of $d\times d$ hermitian matrices $T^a$ 
$(a=1,2,\cdots,d^2-1)$ generating the SU($d$) transformations in 
the fundamental representation, and construct the following operators,
\begin{equation}
Q^a=a_i^\dagger\,T^a_{ij}\,a_j\ \ \ ,\ \ \ {Q^a}^\dagger=Q^a\ \ \ .
\end{equation}
Given the SU($d$) algebra,
\begin{equation}
[T^a,T^b]=if^{abc}T^c\ \ \ ,
\end{equation}
as well as the Jacobi identity for the structure coefficients $f^{abc}$,
it is possible to show that the quantities $Q^a$ 
$(a=1,2,\cdots,d^2-1)$ do indeed generate the same SU($d$) algebra, with the
creation and annihilation operators transforming in the fundamental
representation,
\begin{equation}
[Q^a,Q^b]=if^{abc}Q^c\ \ ,\ \ 
[Q^a,a_i^\dagger]=a_j^\dagger\,T^a_{ji}\ \ ,\ \ 
[Q^a,a_i]=-T^a_{ij}a_j\ \ .
\end{equation}

Consider then finite SU($d$) transformations acting on these operators,
\begin{equation}
e^{i\theta^aQ^a}\,a_i^\dagger\,e^{-i\theta^aQ^a}=
a_j^\dagger\,U_{ji}(\theta^a)\ \ \ ,
\end{equation}
where the matrices $U_{ij}(\theta^a)$ thus define the finite SU($d$) group
transformations in the fundamental representation,
\begin{equation}
U_{ij}(\theta^a)=\left(e^{i\theta^aT^a}\right)_{ij}\ \ \ .
\end{equation}
It is then straightforward to show that the coherent states as well
do indeed transform in the same fundamental SU($d$) representation,
namely,
\begin{equation}
e^{i\theta^aQ^a}\,|z_i>=|U_{ij}(\theta^a)z_j>\ \ \ ,
\end{equation}
a result which thus generalises that of (\ref{eq:rotation}) in the
case of the global SO(2)=U(1) symmetry of the two dimensional
spherical harmonic oscillator.

Finally, it should also be clear that the whole degeneracy of the
quantum Hamiltonian $\hat{H}$ in (\ref{eq:Hd}) at a given energy
level of excitation $\hat{N}=n$, is directly accounted for in terms
of the $n$ times totally symmetric SU($d$) representation, of dimension
$(d-1+n)!/\left((d-1)!\,n!\right)$, whose set of states is thus
generated through the repeated application of the SU($d$) simple positive
roots on the lowest weight state in that representation.

\section*{Appendix B}

Consider the following Euler angle parametrisation of
SO(3) transformations in the defining vector representation,
\begin{equation}
U(\psi,\theta,\phi)=
\left(\begin{array}{c c c}
	\cos\psi & \sin\psi & 0 \\
	-\sin\psi & \cos\psi & 0 \\
	0 & 0 & 1
	\end{array}\right)
\left(\begin{array}{c c c}
	1 & 0 & 0 \\
	0 & \cos\theta & \sin\theta \\
	0 & -\sin\theta & \cos\theta 
	\end{array}\right)
\left(\begin{array}{c c c}
	\cos\phi & \sin\phi & 0 \\
	-\sin\phi & \cos\phi & 0 \\
	0 & 0 & 1
	\end{array}\right)\ \ \ ,
\label{A2:SO(3)matrix}
\end{equation}
where the range of variation for the Euler angles is,
\begin{equation}
0\le\psi\le 2\pi\ \ ,\ \ 
0\le\theta\le \pi\ \ ,\ \ 
0\le\phi\le 2\pi\ \ .
\end{equation}

Introducing the canonical basis of SO(3) generators 
$(T_i)_{jk}=-i\epsilon_{ijk}$, namely,
\begin{equation}
T_1=-i\left(\begin{array}{c c c}
	0 & 0 & 0 \\
	0 & 0 & 1 \\
	0 & -1& 0
	\end{array}\right)\ \ ,\ \ 
T_2=-i\left(\begin{array}{c c c}
	0 & 0 & -1\\
	0 & 0 & 0 \\
	1 & 0 & 0
	\end{array}\right)\ \ ,\ \ 
T_3=-i\left(\begin{array}{c c c}
	0 & 1 & 0 \\
	-1& 0 & 0 \\
	0 & 0 & 0
	\end{array}\right)\ \ ,
\end{equation}
the (left-)invariant Haar integration measure over SO(3) is then
given by the determinant of the matrix defined by considering the
Lie algebra valued differential 
$\left(-iU^{-1}(\psi,\theta,\phi)dU(\psi,\theta\phi)\right)$
expanded both in the Euler angle parametrisation and in the Lie
algebra generators. After some algebra, one then finds that the
normalised Haar measure over SO(3) is given by,
\begin{equation}
\int_{\rm SO(3)}\,d\mu=
\frac{1}{8\pi^2}\int_0^{2\pi}d\psi\,\int_0^\pi d\theta\,\sin\theta\,
\int_0^{2\pi} d\phi\ \ \ .
\end{equation}

Another result of use is the expression for the
determinant ${\rm det}\Big[\one\,-\,x\,U(\psi,\theta,\phi)\Big]$,
where $x$ is some arbitrary parameter. Straightforward algebra leads
to the expression,
\begin{equation}
{\rm det}\Big[\one\,-\,x\,U(\psi,\theta,\phi)\Big]=
(1-x)\left\{(1+x+x^2)+x\left[1-(1+\cos\theta)(1+\cos(\psi+\phi))\right]\right\}
\ \ \ .
\end{equation}

\end{document}